\documentclass[12pt]{article}

\usepackage{amssymb}

\usepackage{graphicx}

\begin{document}

\title{A noncommutative Bianchi I model with radiation}

\author{G. Oliveira-Neto and T. M. Abreu\\
Departamento de F\'{\i}sica, \\
Instituto de Ci\^{e}ncias Exatas, \\ 
Universidade Federal de Juiz de Fora,\\
CEP 36036-330 - Juiz de Fora, MG, Brazil.\\
gilneto@fisica.ufjf.br, thiago\_moralles@ufjf.br}

\maketitle

\begin{abstract}
In the present work, we study the dynamical evolution of an homogeneous and anisotropic, noncommutative (NC) Bianchi I (BI) model coupled to a radiation perfect fluid. Our first motivation is determining if the present model tends to an homogeneous and isotropic NC Friedmann-Robertson-Walker (FRW) model, during its evolution. In order to simplify our task, we use the Misner parametrization of the BI metric. In terms of that parametrization the BI metric has three metric functions: the scale factor $a(t)$ and the two parameters $\beta_\pm (t)$, which measure the spatial anisotropy of the model. Our second motivation is trying to describe the present accelerated expansion of the universe using noncommutativity (NCTY). The NCTY is introduced by two nontrivial Poisson brackets between some geometrical as well as matter variables of the model. We recover the description in terms of commutative variables by introducing some variables transformations that depend on the NC parameter. Using those variables transformations, we rewrite the total NC Hamiltonian of the model in terms of commutative variables. From the resulting Hamiltonian, we obtain the dynamical equations for a generic perfect fluid. In order to solve these equations, we restrict our attention to a model where the perfect fluid is radiation. We solve, numerically, these equations and compare the NC solutions to the corresponding commutative ones. The comparison shows that the NC model may be considered as a possible candidate for describing the accelerated expansion of the universe. Finally, we obtain estimates for the NC parameter and compare the main results of the NC BI model coupled to radiation with the same NC BI model coupled to other perfect fluids. As our main result, we show that the solutions, after some time, produce an isotropic universe. Based on that result, we can speculate that the solutions may represent an initial anisotropic stage of our Universe, that later, due to the expansion, became isotropic.
\end{abstract}



\section{Introduction}

One of the most important areas of study in cosmology is the initial moments of our Universe. Many physicists speculate that during that time
our Universe was very different from what it is today. Some say that during this period the Universe could not be described by general relativity, because the gravitational interaction was so strong that quantum effects were dominant. Therefore, one has to use quantum gravity
theories in order to describe our Universe at such early age. From those quantum gravity theories, we learn that the fundamental concept of
metric tensor is no longer important because it may fluctuate due to quantum uncertainties. Even more radical transformations could take place due to topology changes \cite{wheeler}. Therefore, the ancient Universe could have been very different from the homogeneous and isotropic Universe we know today. But of course, whatever the initial properties it had, during its evolution it reached the present configuration. In the present work, we want to propose an initial configuration of our Universe that is different from the homogeneous and isotropic one. Let us suppose that, initially, the Universe had a well defined metric which was homogenous but anisotropic. More precisely, it was a BI metric. The BI metric was first studied by the Italian mathematician L. Bianchi, who classified the nine three dimensional, homogeneous and anisotropic spaces \cite{bianchi}. The first solution to the Einstein's equations for a cosmological model where the spacetime has the BI metric, was obtained by E. Kasner \cite{kasner}. The BI spacetime is homogeneous and anisotropic and each spatial direction expands (or contracts) in different ways. Therefore, in the usual parametrization it has three scale factors. The three-dimensional spatial sections of this spacetime have zero curvature. As we mentioned, above, in order that our proposal for the initial configuration of our
Universe be acceptable, during its evolution that initial configuration must tend to a homogeneous and isotropic spacetime. This process is  called isotropization. More exactly, it must tend to a FRW spacetime. Due to the homogeneity and isotropy of the FRW spacetime all spacial directions expands (or contracts) in the same way. Then, it has only one scale factor. The FRW spacetime may have spatial sections which have positive, zero or negative curvature. Presently, due to observations, many physicists believe that our present Universe is better described by a FRW spacetime with spatial sections which have zero curvature \cite{george,anton}. Taking this important result in account, our choice of a BI spacetime, in order to describe the initial configuration of our Universe, is very suitable because the spatial sections of this spacetime have already zero curvature. Another important ingredient of our proposal is the matter content of the model. Since, it is believed that the radiation perfect fluid is the most appropriate perfect fluid in order to describe the matter content of the Universe during its early age, we make this choice here. In the literature, one may find several works studying different BI cosmological models where the gravitational theory is general relativity, at the classical or quantum levels. We may mention the following few examples of those models \cite{jacobs,jacobs1,banerjee,biesiada,chakraborty,eath,alvarenga,pradhan,balakin,balakin1,yadav,canfora,kamenshchik,kamenshchik1,mandal,parnovsky,casadio}.

For several years, due to theoretical results in relativistic cosmology, it was accepted that the Universe could end in one of the three following possibilities. In the first one, the Universe would start expanding from an initial big bang singularity until reaches a maximum size, then, it would contract until reaches a big crunch singularity. In the second one, the Universe would start expanding from an initial big bang singularity and would continue to expand, but at a continually decelerating rate, until it stops after an infinity amount of time. Finally, in the third possible end, the Universe would start expanding from an initial big bang singularity and would continue to expand forever with an almost constant expansion rate \cite{MTW}. But in 1998, after the observation of distant supernovas, it was established that the Universe is expanding in an accelerated rate \cite{riess1998}, \cite{perlmutter1999}. There have been many different proposal in order to explain that surprising result. In one proposal, one tries to explain the accelerated expansion due to a new kind of matter with very different properties from the usual matter. For a review on some of those different types of matter see Ref. \cite{Mli}. In another proposal, one tries to explain the accelerated expansion due to modifications in the gravitational sector, more precisely, introducing new gravitational theories other than general relativity. For a review on some ideas belonging to this proposal see Ref. \cite{trodden}. There is, yet, another proposal, where one tries to explain the accelerated expansion of the Universe by deforming the algebra of some variables, that is, by introducing NCTY \cite{vakili,el,obregon,gil,vakili1,sabido,gil1,gil2,pourhassan,gil3,gil4,gil5}.
In the present work, we consider the presence of NCTY, in order to explain the accelerated expansion of the Universe. In particular, we want to investigate if the NCTY may alter the isotropization of the model. The NCTY that we consider, here, are deformations of the Poisson brackets algebra between few metric variables and the fluid variables. In the literature, one may find several examples of the
type of NCTY we use here \cite{vakili,el,obregon,gil,vakili1,sabido,gil1,gil2,pourhassan,gil3,gil4,gil5,garcia,nelson,barbosa,sabido1,gil6,moniz,saba,ghosh,ghosh1,bodmann,barron}. 

In the present work, we study the dynamical evolution of an homogeneous and anisotropic, NC BI model coupled to a radiation perfect fluid. We consider general relativity as the gravitational theory. Our first motivation is determining if the present model tends to an homogeneous and isotropic NC FRW model, during its evolution. More precisely, we want to establish how the parameters and initial conditions of the model, quantitatively, influence the isotropization of the present model. In order to simplify our task, we use the Misner parametrization of the BI metric. In terms of that parametrization the BI metric has three metric functions: the isotropic scale factor $a(t)$ and two parameters $\beta_\pm (t)$, which measure the spatial anisotropy of the model (we shall call them anisotropic scale factors). Our second motivation is trying to describe the present accelerated expansion of the universe using NCTY. We start by writing the total Hamiltonian of the model, which is a sum of the gravitational Hamiltonian with the matter Hamiltonian. Initially, we consider the matter Hamiltonian for a generic perfect fluid. The NCTY is introduced by two nontrivial Poisson brackets between some geometrical as well as matter variables of the model. We associate the same NC parameter $\gamma$ to both nontrivial Poisson brackets. We recover the description in terms of commutative variables by introducing some variables transformations that depend on the NC parameter. Using those variables transformations, we rewrite the total NC Hamiltonian of the model in terms of commutative variables. From the resulting Hamiltonian, we obtain the dynamical equations for a generic perfect fluid. In order to solve these equations, we restrict our attention to a model where the perfect fluid is radiation. Using the solutions to those equations, we compute the NC isotropic scale factor ($a_{nc}(t)$) and we, also, compute the NC anisotropic scale factors ($\beta_{{\pm}_{nc}}(t)$). They all depend on four parameters: $\gamma$, a parameter associated with the fluid energy ($C$) and two parameters which have no clear physical interpretation ($C_\pm$). They also depend on the initial conditions of the model variables. We compare the NC solutions to the corresponding commutative ones. The comparison shows that the NC model may be considered as a possible candidate for describing the accelerated expansion of the universe. Finally, we obtain estimates for the NC parameter and compare some results of the NC BI model coupled to radiation with the same NC BI model coupled to other perfect fluids. As our main result, we show that the solutions, after some time, produce an isotropic universe. Based on that result, we can speculate that the solutions may represent an initial anisotropic stage of our Universe, that later, due to the expansion, became isotropic.

In Section \ref{Bianchi C}, we present the commutative BI model for a generic perfect fluid and obtain the total Hamiltonian. In order to write the fluid Hamiltonian, we use the Schutz variational formalism. In Section \ref{Bianchi NC}, we introduce the NC BI model for a generic perfect fluid, which is characterized by a NC parameter $\gamma$. We obtain the coupled system of ordinary differential equations for the metric functions and the variable associated to the fluid. In Section \ref{NCradiation}, we restrict attention to the case of a radiation perfect fluid. We solve, numerically, the system of ordinary differential equations and obtain the NC scale factors as a function of the time coordinate, the radiation energy density $C$, the anisotropy parameters $C_+$ and $C_-$ and the NC parameter $\gamma$. We vary all the parameters and initial conditions and obtain how they influence the solutions. We compare the model with the commutative case and discuss the differences. In Section \ref{estimativa}, we use some cosmological data in order to estimate the NC parameter. In Section \ref{fluidos}, we compare the NC scale factors behaviors for different perfect fluids. Finally, in Section \ref{conclusoes}, we draw our conclusions.
 
\section{The commutative Bianchi I model for a generic perfect fluid}
\label{Bianchi C}

In the present section, we want to study a cosmological model with the BI metric coupled to a generic perfect fluid.
Since, one of the main objectives is to verify that this model tends, after some time, to a homogeneous and isotropic FRW spacetime, we use the Misner parametrization of the BI metric. It is given by \cite{misner},
\begin{equation}
	ds^2 = - N^2 (t) dt^2 + a^2 (t) \left [e ^{2 (\beta_+ + \sqrt 3 \beta_-)} dx^2 + e ^{2(\beta_+ - \sqrt 3 \beta_-)} dy^2 + e ^{ - 4 \beta_+ } dz^2  \right ],
	\label{newmetric}
\end{equation}
where $a(t)$ is the isotropic scale factor, $\beta_{\pm} (t)$ are the anisotropic scale factors and $N(t)$ is the lapse function coming from the ADM formalism \cite{MTW}. We use the natural unit system where $8\pi G = c = 1$. Observing Eq. (\ref{newmetric}), we notice that if $\beta_+(t)$ and $\beta_-(t)$ tend to zero or constant values, for a given finite or infinity value of time, the Bianchi-I metric Eq. (\ref{newmetric}) goes to the FRW metric, for that given value of time. Therefore, it is very simple to identify the isotropization of the BI spacetime, using the BI metric written in the Misner parametrization Eq. (\ref{newmetric}).

We assume that the universe is filled with a generic perfect fluid, which has an energy-momentum tensor given by,
\begin{equation}
	T_{\mu \nu} = (\rho + p) U_{\mu} U_{\nu} + p g_{\mu \nu}, 
\end{equation}
where $p$ is the fluid pressure, $\rho$ is the fluid energy density and $U^{\mu}$ is the fluid four-velocity, which in a comoving coordinate system has the form, $U^{\mu}= N(t) \delta^{\mu}_{0}$. Furthermore, the equation of state for the perfect fluid is,
\begin{equation}
	p = \omega \rho, 
	\label{eqestado}
\end{equation}
where $\omega$ is a constant which defines the type of fluid.

Here, we adopt the potential-velocity representation from Schutz formalism, where the fluid four-velocity is given by \cite{schutz1970},
\begin{equation}
	U_\nu = \mu^{-1} (\phi_{,\nu} + \alpha \beta_{, \nu} + \theta s_{,\nu}).
	\label{4-velocidade}
\end{equation}
In Eq. (\ref{4-velocidade}), $\mu$ is the specific enthalpy, $s$ is the specific entropy, $\alpha$ and $\beta$ are connected with vorticity and $\phi$ and $\theta$ have no clear meaning. Besides, we consider the four-velocity subjected to the normalization condition,
\begin{equation}
	U^{\nu} U_{\nu} = -1. 
\end{equation}
Furthermore, Schutz formalism provides us with an Eulerian variational principle, in which the Lagrangian density is given by the fluid pressure. So that, the total action, composed by geometry plus matter sectors, is given by,
\begin{equation}
	S = \int d^4 x \sqrt{- g}\left (  R + 2 \, p \right ) ,
	\label{açãoS}
\end{equation}
where $g$ is the determinant of the metric and $R$ is the scalar curvature. 

Since in the BI model there is no vorticity, the potentials $\alpha$ and $\beta$ are zero, so that Eq. (\ref{4-velocidade}) reduces to the following form in a comoving coordinate system with the fluid,
\begin{equation}
	\mu = {N}^{-1} (\dot{\phi} + \theta \dot{s}) \, .
	\label{entalpia}
\end{equation}
For the equation of state Eq. (\ref{eqestado}) it can be shown, through thermodynamics considerations, that the pressure in this formalism takes the form,	
\begin{equation}	
	p = \omega {\Bigl(\frac{\mu}{\omega + 1} \Bigr)^{1 + 1/{\omega}}}\mathrm{e}^{-s/{\omega}}.
	\label{equação de estado}
\end{equation}
From the BI metric Eq. (\ref{newmetric}) it is not difficult to find that $\sqrt{ - g} = N a^3 $ and the Ricci scalar is,
\begin{equation}
\label{R}
	R = 6 {\frac{\ddot{a}}{N^2 a}} - 6 \frac{\dot{a} \dot{N}}{N^3 a} + 6 \frac{\dot{ a}^2}{N^2 a^2} + 6 \frac{\dot{\beta_+}^2}{N^2} + 6 \frac{\dot{\beta_-}^2}{N ^2}.
\end{equation}
Thus, the total action Eq. (\ref{açãoS}) for BI can be cast in the form, with the aid of Eqs. (\ref{entalpia})-(\ref{R}),
\begin{equation}
\label{SL}
S = \int \mathcal{L} dt,
\end{equation}
where,
\begin{equation}
\label{L}
\mathcal{L} = - 6 \frac{\dot{a}^2 a}{N} + 6 \frac{\dot{a^3 \beta_+}^2}{N} + 6 \frac{a^3 \dot{\beta_-}^2}{N} + \, 2 \, N^{-1/ \omega} \, \frac{a^3 \omega}{(1 + \omega)^{1 + 1/\omega}} \, (\dot{\phi} + \theta \dot{s})^{1 + 1/\omega}
		\, e^{{-s}/{\omega}}  .  
\end{equation}	
The total Hamiltonian of the model is therefore,
\begin{equation}
	N \mathcal{H}  = N \left [ a^{-3} \left (- \frac{ p_{a}^2 a^2}{24} + \frac{ p_{+}^2}{24} + \frac{ p_{-}^2}{24} \right ) + 2 a^{- 3 \omega } e^{s} p_\phi^{1 + \omega} \right ] \, ,
	\label{superhamiltoniana}
\end{equation}
where $ p_{a} = -12a\dot{a}/N$, $ p_{+} = 12a^3 \dot{\beta}_+ /N$, $ p_{-} = 12a^3 \dot{\beta}_- /N$, $ p_\phi = N^{-1/\omega} a^3 (\dot{\phi} + \theta\dot{s})^{1/\omega} \exp{(-s/\omega)}/(\omega + 1)^{1/\omega}$ and $ p_s = \theta p_\phi $ are the momenta canonically conjugate to $a$, $\beta_{+}$, $\beta_{-}$, $\phi$ and $s$, respectively.   

We may simplify the Hamiltonian Eq. (\ref{superhamiltoniana}) by performing the following canonical transformations \cite{lapchinskii1977}, 
\begin{equation}
	T = p_s e^{-s} p_\phi^{- (1 + \omega)},  \,\ p_{T} = 2p_\phi^{1 + \omega} e^s, \,\  \bar{\phi} = \phi - (1 + \omega) \frac{p_s}{p_\phi} , \,\  \bar{p_\phi} = p_\phi ,
\end{equation}
where $T$ is a canonical variable associated to the fluid.  Thus, the total Hamiltonian Eq. (\ref{superhamiltoniana}) becomes,
\begin{equation}
	N_{} \mathcal{H}_{}  = - \frac{{p_{{a}_{}}^2}}{24 a_{}} + \frac{p_{{+}_{}}^2}{24 a^{3}_{}} + \frac{p_{{-}_{}}^2}{24 a^{3}_{}}  + {a_{}}^{- 3 \omega}  p_{{T}_{}} \, ,
	\label{Hcomutativo}
\end{equation}
where $p_{{T}_{}}$ is the momentum canonically conjugate to $T$. In the present work, we choose the cosmological gauge, where $N = 1$.

\section{The noncommutative Bianchi I model for a generic perfect fluid}
\label{Bianchi NC}

In our model we will consider that the total Hamiltonian of the NC model has the same functional form as the commutative model, which is given by Eq. (\ref{Hcomutativo}). Thus, we write it in terms of NC variables as,
\begin{equation}
	N_{nc} \mathcal{H}_{nc} = - \frac{{p_{{a}_{nc}}^2}}{24 a_{nc}} + \frac{p_{{+}_{nc}}^2}{24 a^{3}_{nc}} + \frac{p_{{-}_{nc}}^2}{24 a^{3}_{nc}} 
	 + {a_{nc}}^{- 3 \omega}  p_{{T}_{nc}} \, ,
	\label{NH-NC}
\end{equation}
where we choose the gauge $N_{nc} = 1$.

We impose that the NC variables of the model ($a_{nc}$, $p_{a_{nc}}$, $\beta_{{\pm}_{nc}}$, $p_{{\pm}_{nc}}$, $T_{nc}$, $p_{T_{nc}}$) satisfy the following deformed Poisson brackets (PBs),
\begin{equation}
	\{ a_{nc}, p_{T_{nc}} \} =	
	\{ T_{nc}, p_{a_{nc}} \} = \gamma \, , \,\
	\label{PBs}
\end{equation}
in addition, of course, to the usual PBs,
\begin{equation}
		\{ {a_{nc}, p_{a_{nc}}} \} =   
	\{ \beta_{{\pm}_{nc}}, p_{{\pm}_{nc}} \} =   
	\{ T_{nc}, p_{T_{nc}} \} = 1 \, ,
\label{PBs1}
\end{equation}
and with all other combinations equal to zero. The term $\gamma$ in Eq. (\ref{PBs}) is the NC parameter. As we are considering that the NCTY is only a small residual effect nowadays, the NC parameter is very small ($\left | \gamma \right | \ll 1 $). Therefore, in all calculations and expressions with the NC parameter it will be considered up to the first order, unless otherwise indicated.

To simplify our description of the NC model we can introduce canonical transformations connecting the NC variables ($a_{nc}$, $p_{a_{nc}}$, $\beta_{{\pm}_{nc}}$, $p_ {{\pm}_{nc}}$, $T_{nc}$, $p_{T_{nc}}$) with new commutative ones ($a_{c}$, $p_{a_{c}}$, $\beta_{{\pm}_{c}}$, $p_{{\pm}_{c}}$, $T_{c}$, $p_{T_{c}}$). These new commutative variables must satisfy the usual PBs instead of the deformed PBs. Taking into account the PBs in Eq. (\ref{PBs}), the following transformations lead from the NC variables to the commutative ones,
\begin{eqnarray}
	\label{parentes1}
	& & a_{nc} \rightarrow a_{c} + \frac{\gamma}{2} T_{c} \, , \,\ p_{a_{nc}} \rightarrow p_{a_{c}} + \frac{\gamma}{2} p_{T_{c}} \, , \\
	\label{parentes2}
	& & \beta_{{\pm}_{nc}} \rightarrow \beta_{{\pm}_{c}} \, , \,\
	p_{{\pm}_{nc}} \rightarrow p_{{\pm}_{c}} \, , \\
	\label{parentes3}
	& & T_{nc} \rightarrow T_{c} + \frac{\gamma}{2} a_{c} \, , \,\ p_{T_{nc}} \rightarrow p_{T_{c}} + \frac{\gamma}{2} p_{a_{c}} \, , 
\end{eqnarray}	
where the commutative variables have $c$ labels. Furthermore, the NC variables given by Eqs. (\ref{parentes1})-(\ref{parentes3}) satisfy, to first-order in $\gamma$, the PBs Eqs. (\ref{PBs}), (\ref{PBs1}) and the others which are nil, when we use the usual PBs among the commutative variables. It is important to say that such transformations Eqs. (\ref{parentes1})-(\ref{parentes3}) are not unique, and others could also be proposed. That said, we can rewrite the total NC Hamiltonian, $N_{nc} \mathcal{H}_{nc}$ Eq. (\ref{NH-NC}) in terms of the commutative variables Eqs. (\ref{parentes1})-(\ref{parentes3}), so that,
\begin{eqnarray}
	\mathcal{H}_{nc} & = & 
	- \frac{ {(p_{a_{c}} + \frac{\gamma}{2} p_{T_{c}})^2}}{24 (a_{c} + \frac{\gamma}{2} T_{c})} + \frac{ p_{{+}_{c}}^2}{24 (a_{c} + \frac{\gamma}{2} T_{c})^3} + \frac{ p_{{-}_{c}}^2}{24 (a_{c} + \frac{\gamma}{2} T_{c})^3}\nonumber \\
		 & + & \left(a_{c} + \frac{\gamma}{2} T_{c}\right)^{- 1} \left(p_{T_{c}} + \frac{\gamma}{2} p_{a_{c}}\right) \, .
	\label{H total}
\end{eqnarray} 

From $\mathcal{H}_{nc}$ Eq. (\ref{H total}), we compute Hamilton's equations with the help of the usual PBs between the commutative variables. This results in the following set of equations,
\begin{equation}
	\dot{a}_{c} = \{a_{c} , \mathcal{H}_{nc} \} = - \frac{1}{12}\frac{(p_{a_{c}} + \frac{\gamma}{2} p_{T_{c}})}{(a_{c} + \frac{\gamma}{2} T_{c})} + \frac{\gamma}{2}\left ( a_{c} + \frac{\gamma}{2} T_{c} \right )^{- 3 \omega} \, ,
	\label{a ponto}
\end{equation}
\begin{eqnarray}
	\dot{p}_{a_{c}} & = & \{p_{a_{c}} , \mathcal{H}_{nc} \} = - \frac{1}{24}\frac{(p_{a_{c}} + \frac{\gamma}{2} p_{T_{c}})^2}{(a_{c} + \frac{\gamma}{2} T_{c})^2} + \frac{1}{8}\frac{p_{+_c}^2}{(a_{c} + \frac{\gamma}{2} T_{c})^4} + \frac{1}{8}\frac{p_{-_c}^2}{(a_{c} + \frac{\gamma}{2} T_{c})^4}\nonumber \\ 
	& + & 3{\omega}\frac{(p_{T_c} + \frac{\gamma}{2} p_a)}{(a_{c} + \frac{\gamma}{2} T_{c})^{3 \omega + 1}}
	\, ,
	\label{p_a ponto}
\end{eqnarray}
\begin{equation}
	\dot{\beta}_{{\pm}_{c}} = \{\beta_{{\pm}_{c}} , \mathcal{H}_{nc}\} =  \frac{1}{12}\frac{p_{{\pm}_{c}}}{(a_{c} + \frac{\gamma}{2} T_{c})^3} \, ,
	\label{betadot}
\end{equation}
\begin{equation}
	\dot{p}_{{\pm}_c} = \{p_{{\pm}_c}, \mathcal{H}_{nc}\} = 0 \, ,	
	\label{p pm ponto}
\end{equation}
\begin{equation}
	\dot{T}_c = \{T_c, \mathcal{H}_{nc}\} = -\frac{\gamma}{24} \frac{(p_{a_{c}} + \frac{\gamma}{2} p_{T_{c}})}{(a_{c} + \frac{\gamma}{2} T_{c})} + \left (a_{c} + \frac{\gamma}{2} T_{c} \right )^{-3 \omega} \, ,	
	\label{T ponto}
\end{equation}
\begin{eqnarray}
	\dot p_{T_{c}} & = & \{p_{T_{c}}, \mathcal{H}_{nc}\} = \frac{\gamma}{2} [ -\frac{1}{24} \frac{(p_{a_{c}} + \frac{\gamma}{2} p_{T_{c}})^2}{(a_{c} + \frac{\gamma}{2} T_{c})^2} + \frac{1}{8}\frac{p_{+_c}^2}{(a_{c} + \frac{\gamma}{2} T_{c})^4} + \frac{1}{8}\frac{p_{-_c}^2}{(a_{c} + \frac{\gamma}{2} T_{c})^4}\nonumber \\ 
	& + & {3 \omega} \frac{(p_{T_c} + \frac{\gamma}{2} p_a)}{(a_{c} + \frac{\gamma}{2} T_{c})^{3 \omega + 1}} ] \, .
	\label{p_T ponto}
\end{eqnarray}

With the objective of determining the behavior of the commutative isotropic scale factor and the commutative anisotropic scale factors, we combine Eqs. (\ref{a ponto})-(\ref{p_T ponto}). It allows us to write a set of four differential equations for $a_c ( t)$, $\beta_{+_{c}} (t)$, $\beta_{-_{c}} (t)$, $T_c (t)$ and their time derivatives. Starting by combining Eqs. (\ref{p_a ponto}) and (\ref{p_T ponto}) we find,
\begin{equation}
	p_{T_c} = \frac{\gamma}{2}p_{a_c} + C \, ,
	\label{pT pa C}
\end{equation}
where $C$ is an integration constant. In physical terms, for the commutative case ($\gamma = 0$), $C$ represents the energy density of the fluid at a certain time, therefore a positive quantity. Furthermore, integrating Eqs. (\ref{p pm ponto}), the solutions can be conveniently put in the form,
\begin{equation}
	p_{\pm} = \sqrt{C_{\pm}},
	\label{p pm}
\end{equation}
where $C_+$ and $C_-$ are constants that do not have a clear physical meaning, but are connected to the degrees of freedom of the anisotropies. 

Now, combining the Eqs. (\ref{a ponto}) and (\ref{T ponto}), we obtain the following equation for the temporal evolution of the variable $T(t)$,
\begin{equation}
	\dot{T_{c}} = \frac{\gamma}{2}{\dot{a}_{c}} + \left ( a_{c} + \frac{\gamma}{2}T_{c } \right )^{- 3 \omega} \, .
	\label{T ponto1}
\end{equation}

On the other hand, from Eqs. (\ref{a ponto}) and (\ref{pT pa C}) it is possible to find an expression for $p_{a_{c}}$,
\begin{equation}
	p_{a_{c}} = - 12 \dot{a_{c}}a_{c} - 6 \gamma \dot{a_{c}} T_{c} + 6 \gamma a_{c} \left ( a_ {c} + \frac{\gamma}{2}T_{c} \right )^{- 3 \omega} -\frac{\gamma}{2} C \, .
	\label{p a c}
\end{equation}

The acceleration equation for the isotropic scale factor can be found by deriving, with respect to $t$, Eq. (\ref{a ponto}) and substituting in the resulting expression, Eqs. (\ref{p_a ponto}), (\ref{T ponto}), (\ref{pT pa C}) and (\ref{p a c}). The calculation, quite extensive but straightforward, results in,
\begin{eqnarray}
	\ddot{a}_{c} & = & - \frac{1}{2} {\left ( a_{c} + \frac{\gamma}{2}T_{c} \right )}^{-1} [ \dot{a}^2_{c} + \frac{(1 - 3 \omega) \dot{a}_{c} \gamma }{\left ( a_{c} + \frac{\gamma }{2}T_{c} \right )^{3 \omega}} + \frac{1}{48} \frac{(C_{+} + C_{-})}{\left ( a_{c} + \frac{\gamma}{2}T_{c} \right )^4}\nonumber \\ 
	 & + & \frac{1}{2} \frac{C \omega}{\left ( a_{c} + \frac{\gamma }{2}T_{c} \right )^{3 \omega + 1}} ] \, ,
	\label{evoA}
\end{eqnarray}

Next, taking the time derivative of Eq. (\ref{betadot}) and using Eq. (\ref{p pm}), we obtain,
\begin{equation}
	\ddot{{\beta}}_{{\pm}_{c}} = - \frac{1}{4} \frac{\left ( \dot{a}_c + \frac{\gamma}{2 } \dot{T}_c \right )}{\left ( a_c + \frac{\gamma}{2} T_c \right )^4} \, \sqrt{C_{\pm}} \, .
	\label{evoa}
\end{equation}
These expressions Eqs. (\ref{evoa}), can be rewritten in a convenient way, if one uses Eqs. (\ref{betadot}) again. The result are two identical equations for the anisotropic scale factors, given by,
\begin{equation}
	{3 H}{ \left ( a_{c} + \frac{\gamma}{2}T_{c} \right )}^{-1} {\left [ a_{c} + \frac{\gamma} {2 H} \left ( a_{c} + \frac{\gamma}{2}T_{c} \right )^{- 3 \omega}\right ]} \dot{\beta}_{\pm_c} + \ddot{\beta}_{\pm_c} = 0 \, ,
	\label{evoBeta}
\end{equation}
where $H = \dot{a}_{c} / {a}_{c}$ is the Hubble parameter for the commutative case.
The Eqs. (\ref{evoBeta}) provide the evolution of the anisotropic scale factors. 
Therefore, we have, now, a system of four ordinary differential equations (\ref{T ponto1}), (\ref{evoA}), (\ref{evoBeta}) in order to determine the dynamics of the four variables $( a_c(t), \beta_{\pm_c} (t), T(t) )$. With the objective of solving these equations and computing the value of all variables, we must provide initial conditions for $a_{c} (t)$, $\dot{a}_{c} (t)$, $\beta_{\pm_c} (t)$, $\dot{\beta}_{\pm_c} (t)$ and $T_{ c} (t)$. By setting $\gamma = 0$ in Eqs. (\ref{T ponto1}), (\ref{evoA}), (\ref{evoBeta}), the evolution equations for the commutative BI model are recovered, in the given parametrization, for $N=1$.

In addition to the system of four equations represented by Eqs. (\ref{T ponto1}), (\ref{evoA}) and (\ref{evoBeta}), the model is subject to constraints, which delimit the physically acceptable initial conditions. These constraints are first order differential equations in the variables of our model. The first one can be computed by imposing the superhamiltonian constraint $\mathcal{H}_{nc} = 0$. So, doing this in Eq. (\ref{H total}) and using Eqs. (\ref{pT pa C}), (\ref{p pm}) and (\ref{p a c}), one gets,
\begin{eqnarray}
&	 & \frac{1}{2} \frac{\dot{a}^2_{c}a_{c}^2}{\left ( a_{c} + \frac{\gamma}{2}T_{c} \right )} + \frac{1}{2} {\dot{a}^2_{c} \gamma T_{c}} + \frac{1}{2}\frac{\dot{a}_{ c} \gamma}{\left ( a_{c} + \frac{\gamma}{2}T_{c} \right )^{3 \omega - 1}} - \frac{1}{288}\frac {(C_+ + C_-)}{\left ( a_{c} + \frac{\gamma}{2}T_{c} \right )^3}\nonumber \\
	 & - & \frac{1}{12} \frac{C }{\left ( a_{c} + \frac{\gamma}{2}T_{c} \right )^{ 3 \omega}} = 0 \, .
	\label{tipo friedmann}
\end{eqnarray}
Furthermore, two other constraint equations emerge from considering the combination of Eqs. (\ref{betadot}) and (\ref{p pm}), from which one obtains,
\begin{equation}
	\dot{\beta}_{{\pm}_{c}} = \frac{1}{12} \frac{\sqrt{C_{\pm}}}{(a_{c} + \frac{\gamma}{2} T_{c})^3} .
	\label{evoluçãobetas}
\end{equation}

Therefore, we obtain a system of equations given by Eqs. (\ref{T ponto1}), (\ref{evoA}) and (\ref{evoBeta}) subject to the constraints 
Eqs. (\ref{tipo friedmann}) and (\ref{evoluçãobetas}) for which we must look for solutions. Nevertheless, it should be emphasized that for the present NC model the physical variables are the scale factors $a_{nc} (t)$ and $\beta_{{\pm}_{nc}} (t)$, which are given by,
\begin{equation}
	 a_{nc} (t) = a_{c} (t) + \frac{\gamma}{2}  T_{c} (t),
	\label{anc1}
\end{equation}
\begin{equation}
	\beta_{{\pm}_{nc}} (t) = \beta_{{\pm}_{c}} (t) , 
	\label{leiBetas}
\end{equation}
these quantities come from Eqs. (\ref{parentes1}) and (\ref{parentes2}). 
It means that, after solving the system of Eqs. (\ref{T ponto1}), (\ref{evoA}) and (\ref{evoBeta}) subject to the constraints 
Eqs. (\ref{tipo friedmann}) and (\ref{evoluçãobetas}), we must combine the commutative variables following Eqs. (\ref{anc1}) and (\ref{leiBetas}) in order to determine the physical scale factors of the model.
In the next section, we will evaluate these quantities for a particular perfect fluid.

\section{The noncommutative Bianchi-I model for a radiation perfect fluid}
\label{NCradiation}

The system of equations given by Eqs. (\ref{T ponto1}), (\ref{evoA}) and (\ref{evoBeta}), depends on the value of $\omega$, that is, on the choice of the perfect fluid that represents the matter content of the universe. Here, we will choose the value $\omega = 1/3$, which represents a radiation perfect fluid. Such a choice is quite reasonable to describe the early universe, when this type of matter was very important \cite{MTW}. Therefore, we will investigate how the BI model with radiation behaves with the inclusion of NCTY. With this choice the system of equations (\ref{T ponto1}), (\ref{evoA}) and (\ref{evoBeta}) simplify to,
\begin{equation}
	\dot{T_{c}} = \frac{\gamma}{2}{\dot{a}_{c}} + \left ( a_{c} + \frac{\gamma}{2}T_{c} \right )^{- 1} \, ,
	\label{T ponto R}
\end{equation}
\begin{equation}
	\ddot{a}_{c} = - \frac{1}{2} {\left ( a_{c} + \frac{\gamma}{2}T_{c} \right )}^{-1} \left [ \dot{a}^2_{c} +  \frac{1}{48} \frac{(C_{+} + C_{-})}{\left ( a_{c} + \frac{\gamma}{2}T_{c} \right )^4} + \frac{1}{6} \frac{C}{\left ( a_{c} + \frac{\gamma}{2}T_{c} \right )^{2}} \right ] \, ,
	\label{evoaR}
\end{equation}
\begin{equation}
	{3 H}{ \left ( a_{c} + \frac{\gamma}{2}T_{c} \right )}^{-1}  {\left [ a_{c} + \frac{\gamma}{2 H} \left ( a_{c} + \frac{\gamma}{2}T_{c} \right )^{- 1}\right ]} \dot{\beta}_{\pm_c} + \ddot{\beta}_{\pm_c} = 0 \, , 
	\label{evoBetaR}
\end{equation}
while the constraint equation (\ref{tipo friedmann}) becomes,
\begin{equation}
	\frac{1}{2} \frac{\dot{a}^2_{c}a_{c}^2}{\left ( a_{c} + \frac{\gamma}{2}T_{c} \right )} + \frac{1}{2} {\dot{a}^2_{c} \gamma (T_{c} + 1)}  - \frac{1}{288}\frac {(C_+ + C_-)}{\left ( a_{c} + \frac{\gamma}{2}T_{c} \right )^3} - \frac{1}{12} \frac{C }{\left ( a_{c} + \frac{\gamma}{2}T_{c} \right )} = 0 \, .
	\label{tipo friedmann R}
\end{equation}
The other two constraint equations (\ref{evoluçãobetas}) are not altered.
So, we will have to solve the system of four equations (\ref{T ponto R}), (\ref{evoaR}) and (\ref{evoBetaR}) subject to the three constraints (\ref{evoluçãobetas}) and (\ref{tipo friedmann R}). Unfortunately, we could not find algebraic solutions to this system of equations. Therefore, we solve it numerically. In order to investigate how the solutions, to the system of equations, depend on each parameter $\gamma$, $C$, $C_{\pm}$ and on each initial conditions $a_c (0)$, $\dot{a}_{c} (0)$, $\beta_{\pm} (0)$ and $\dot{\beta}_{\pm} (0)$, we vary that parameter or initial condition while keeping all others fixed. As a matter of simplicity, we are not going to study how the solutions depend on the initial condition $T_{c} (0)$. It will be fixed at the value $T_{c} (0) = 0$. We carry out this study for a large number of different values of each parameter and initial condition. Note that, we are abandoning the subscript $c$ below $\beta_{\pm}$, by virtue of Eq. (\ref{leiBetas}). We adopt the premise that whenever we are not studying the initial conditions $a_c (0)$ and $\beta_{\pm} (0)$, they assume the following values,
\begin{equation}
 a_c (0) = 1 ,
\end{equation}
\begin{equation}
\beta_{\pm} (0) = 1 .
\end{equation}
When varying the initial conditions, we consider all of them as positive. In order to find physically consistent initial conditions, when we 
vary a parameter or an initial condition, we must fix all other quantities and solve the constraint equations (\ref{evoluçãobetas}) and 
(\ref{tipo friedmann R}). We solve it for a chosen parameter or initial condition that must be let free to vary along with the parameter or initial condition that we are studying. For the study of all parameters and initial conditions, with the exception of $\dot{a}_{c} (0)$,
we let that initial condition varies. When we study $\dot{a}_{c} (0)$, we let $C$ varies and when we study $\dot{\beta}_{\pm} (0)$, we let $C_{\pm}$ vary.

To study how the scale factors depend on each parameter or initial condition of the model, we plotted two types of graphs for each parameter or initial condition investigated. One is the graph $a_{nc} \times t$ and the other is the graph $\beta_{+} \times t$. The behavior of $\beta_{-}$ is qualitatively similar to that of $\beta_{+}$, being identical when $C_{+} = C_{-}$, $\beta_{+} (0) = \beta_{-} (0)$ and $\dot{\beta}_{+} (0) = \dot{\beta}_{-} (0)$, due to Eqs. (\ref{evoluçãobetas}) and (\ref{evoBetaR}). For simplicity, we make this choice and present only the graphical behavior of $\beta_{+}$, from which $\beta_{-}$ follows analogously. It is important to say that we obtained many solutions using different values of the parameters and initial conditions. The graphs that we are going to present are examples which show the general behavior of the solutions in a very clear way. Since we are interested in studying the isotropization of the model, we restrict our attention to expansive solutions of $a_{nc}$.

\subsection{\textbf{$\gamma$ variation}}

Now, we want to investigate how the solutions to the system of Eqs. (\ref{T ponto R}), (\ref{evoaR}) and (\ref{evoBetaR}) depend on the
NC parameter $\gamma$. Then, varying it and keeping all other parameters and initial conditions fixed, it was found that $a_{nc}$ expands more rapidly when we increase the modulus of $\gamma$. If we take
two different values of $\gamma$ with the same modulus but opposite signs, $a_{nc}$ expands more rapidly for the negative value of $\gamma$. Therefore, the introduction of NCTY produces the effect of increasing the scale factor expansion rate. In Figure 1, we present an example of that effect, in a graph of $a_{nc} \times t$, for five different values of the NC parameter ($\gamma$ = -0.5, -0.25, 0, 0.25 and 0.5), with the value $\gamma$ = 0 corresponding to the commutative case. Next, studying how the variation of $\gamma$ influences the dynamics of
$\beta_\pm$, we notice that for all values of $\gamma$, $\beta_\pm$ always goes to a constant value after a period of expansion. 
The greater the modulus of $\gamma$ the more rapidly $\beta_\pm$ goes to a constant value. If we take two different values of $\gamma$ with the same modulus but opposite signs, $\beta_\pm$ go to a constant value more rapidly for the negative value of $\gamma$.
In Figure 2, we present an example of that effect, in a graph of $\beta_{+} \times t$ (as explained, above, $\beta_{-}$ behaves in the same way), where we use the same five values of $\gamma$ given in Figure 1. In Table \ref{T1}, we present the values of the scale factors and their time derivatives for different times and $\gamma$ values. It is noted that the values of $\beta_{+} (t)$ stabilize while $\dot{\beta}_{+} (t)$ tend to zero, which characterizes the phenomenon of isotropization. On the other hand, $a_{nc} (t)$ is always expansive, although in a decelerated way, as shown by the values of $\dot{a}_{nc} (t)$.

	\begin{figure}[!tbp]
	\begin{minipage}{0.4\textwidth}
		\centering
		\label{fig1}
		\includegraphics[width=\linewidth]{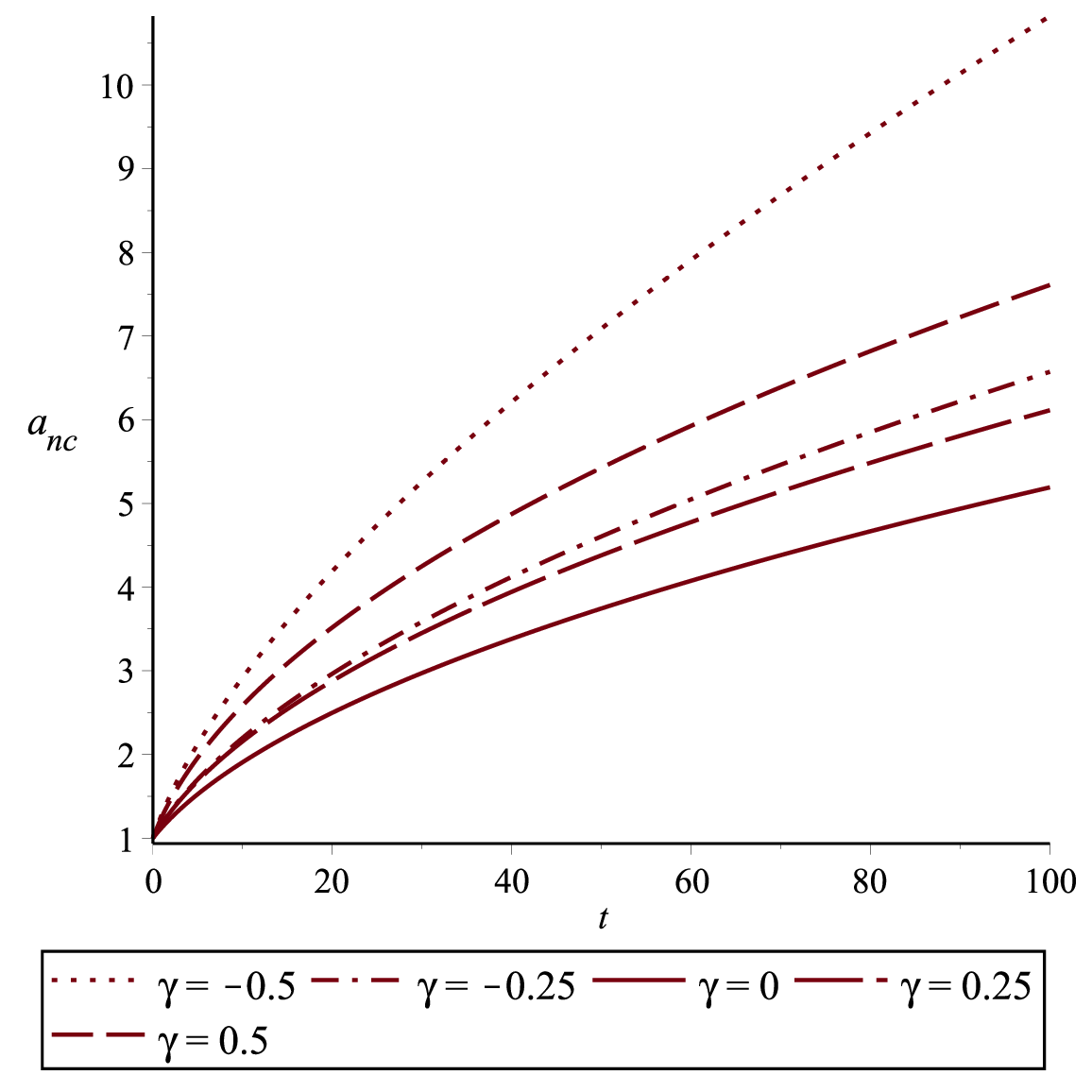}
		{\small Figure 1: $a_{nc} (t) \times t$ for different values of $\gamma$. Here, we considered $C$ = $C_+$ = $C_-$ = 0.1 and $\dot{\beta}_{\pm} (0) = 0.026352314$.}
	\end{minipage}\hfill
	\begin{minipage}{0.4\textwidth}
		\centering
		\label{fig2}
		\includegraphics[width=\linewidth]{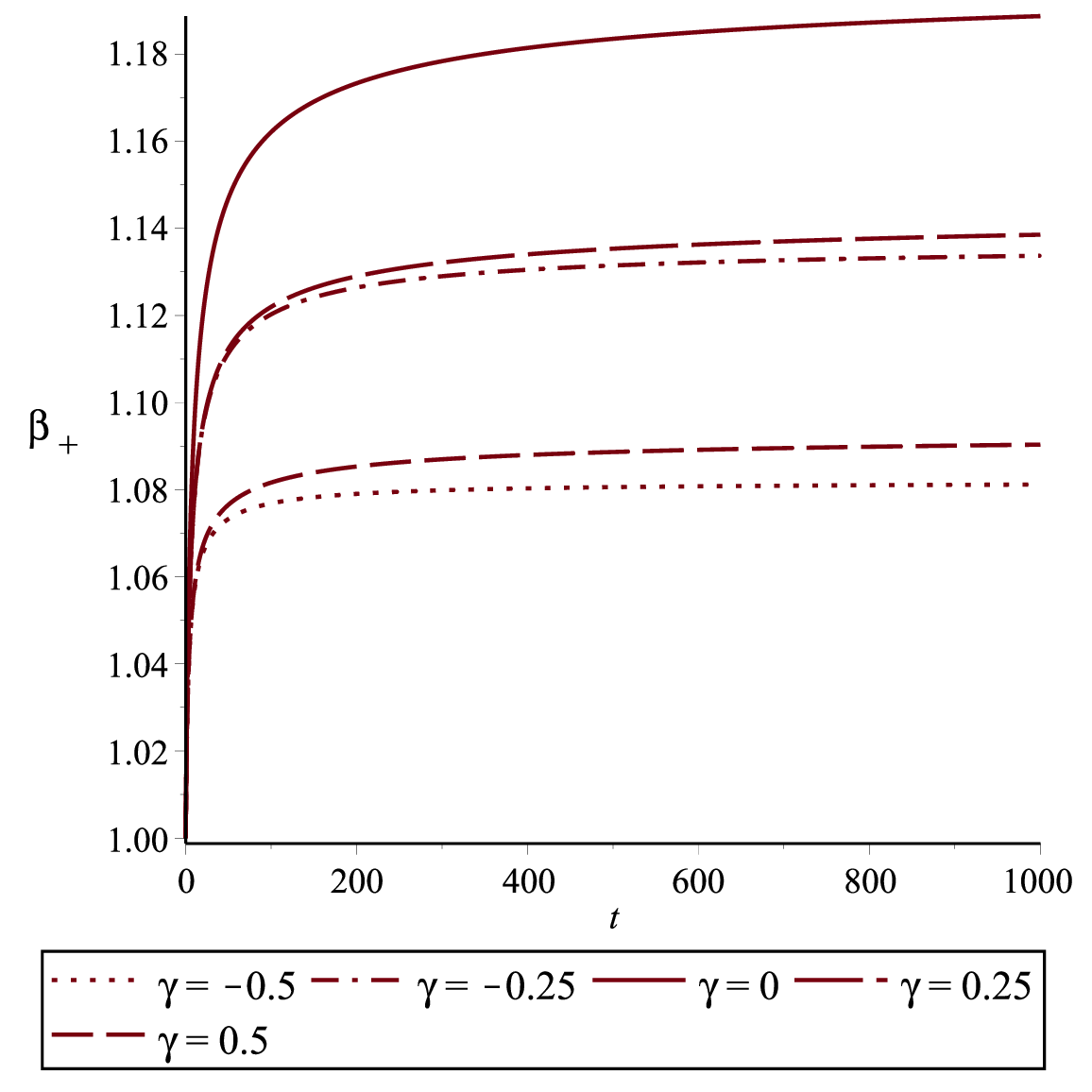}
		{\small Figure 2: $\beta_{+} (t) \times t$ for different values of $\gamma$. Here, we considered $C$ = $C_+$ = $C_-$ = 0.1 and $\dot{\beta}_{\pm} (0) = 0.026352314$.}
	\end{minipage}\hfill
\end{figure}

\begin{table}[!tbp]
	\centering
	\caption{\small Values of $a_{nc}$, $\dot{a}_{nc}$, $\beta_+$ and $\dot{\beta}_{+}$ for different values of $\gamma$ and $t$, with $C = C_+ = C_- = 0.1$ and $\dot{\beta}_{\pm} (0) = 0.026352314$.}
	\label{T1}
		{\tiny\begin{tabular}{c c c c c c}
				\hline
				$\gamma$ &  $t $ & $a_{nc} (t)$ & $\dot{a}_{nc} (t)$ & $\beta_{+} (t)$ & $\dot{\beta}_{+} (t)$ \\
				\hline
				
				-0.5 & $10^{3}$ & $5.77494479470455 \times 10$ & $3.52018296425835 \times 10^{-2}$ & $1.08116411704404$ & $6.28006743211983 \times 10^{-7}$ \\
				
				-0.5 & $10^{5}$ & $1.09169175591580 \times 10^{3}$ & $7.20701293489404 \times 10^{-3}$ & $1.08186070638515$ & $9.83964209788674 \times 10^{-11}$ \\ 
				
				-0.5 & $10^{10}$ & $2.58968680211751 \times 10^{6}$ & $1.76007813141873  \times 10^{-4}$ & $1.08187141805590$ & $2.48074434412398 \times 10^{-20}$ \\ 
				
				-0.5 & $10^{50}$ & $4.10033296508804 \times 10^{33}$ & $2.78822525402685 \times 10^{-17}$ & $1.08187141832555$ & $3.93827793662011 \times 10^{-97}$ \\    
				
				-0.5 & $10^{100}$ & $4.10027918315254 \times 10^{67}$ & $2.78818974557760 \times 10^{-33}$ & $1.08187141832555$ & $3.94654805491996 \times 10^{-193}$ \\    
				
				\hline
				
				- 0.25 & $10^{3}$ & $3.41742390623054  \times 10$ & $1.87873076592792 \times 10^{-2}$ & $1.13370473164562$ & $2.52057532900416 \times 10^{-6}$ \\
				
				- 0.25 & $10^{5}$ & $4.98206113645293 \times 10^{2}$ & $3.07088173114634 \times 10^{-3}$ & $1.13683618564938$ & $4.94759679595310 \times 10^{-10}$ \\  
				
				- 0.25 & $10^{10}$ & $9.35034362858757 \times 10^{5}$ & $6.25250437547558 \times 10^{-5}$ & $1.13688782725201$ & $6.63693932416572 \times 10^{-20}$ \\ 
				
				- 0.25 & $10^{50}$ & $5.93117627592570 \times 10^{32}$ & $3.97450035862926\times 10^{-18}$ & $1.13688782792970$ & $4.44125363487530 \times 10^{-99}$ \\      
				
				- 0.25 & $10^{100}$ & $1.89797399373580 \times 10^{66}$ & $1.27183803551766 \times 10^{-34}$ & $1.13688782792970$ & $4.77974254997572 \times 10^{-198}$ \\    
				
				\hline
				
				0 & $10^{3}$ & $1.61067275263554 \times 10$ & $8.01647757249714 \times 10^{-3}$ & $1.18871845023854$ & $6.30674162922823 \times 10^{-6}$ \\
				
				0 & $10^{5}$ & $1.60685796904401 \times 10^{2}$ & $8.03364665466973 \times 10^{-4}$ & $1.20012123394029$ & $6.35198079492232 \times 10^{-9}$ \\  
				
				0 & $10^{10}$ & $5.03747740162615 \times 10^{4}$ & $2.49664702465128 \times 10^{-6}$ & $1.20138818988449$ & $2.06175314548412 \times 10^{-16}$ \\
				
				0 & $10^{12}$ & $4.60881374729826 \times 10^{5}$ & $2.04673815569800 \times 10^{-7}$ & $1.20139207001625$ & $2.69227313155151 \times 10^{-19}$ \\    
				
				0 & $10^{13}$ & $9.88747417345317 \times 10^{5}$ & $8.41973405201580 \times 10^{-10}$ & $1.20139258308111$ & $2.72669030625673 \times 10^{-20}$ \\ 
				
				\hline
				
				0.25 & $10^{3}$ & $6.70936033474032$ & $3.11696772538865 \times 10^{-3}$ & $1.13853150480785$ & $3.88019973229339 \times 10^{-6}$\\
				
				0.25 & $10^{5}$ & $8.17978892859367 \times 10$  & $5.14876440917777 \times 10^{-4}$ & $1.14531801756597$ & $3.18822540198597 \times 10^{-9}$ \\  
				
				0.25 & $10^{10}$ & $2.78814114084560 \times 10^{5}$ & $1.90607027888036 \times 10^{-5}$ & $1.14578452135441$ & $1.45821730100333 \times 10^{-18}$ \\    
				
				0.25 & $10^{50}$ & $1.85079239840570 \times 10^{32}$ & $1.24022176541388 \times 10^{-18}$ & $1.14578453626068$ & $9.78459030687123 \times 10^{-98}$ \\   
				
				0.25 & $10^{100}$ & $5.92252759735156 \times 10^{65}$ & $3.96870243043003 \times 10^{-35}$ & $1.14578453626068$ & $1.05301913870114 \times 10^{-196}$ \\  
				
				\hline
				
				0.5 & $10^{3}$ & $3.75045200031356$ & $1.54062195723466 \times 10^{-3}$ & $1.09032992612216$ & $2.04623741709548 \times 10^{-6}$\\
				
				0.5 & $10^{5}$ & $4.93801638613804 \times 10$ & $3.47896875116609 \times 10^{-4}$ & $1.09397656783531$ & $ 1.84993582143270 \times 10^{-9}$ \\   
				
				0.5 & $10^{10}$ & $3.09276879590900 \times 10^{5}$ & $2.16494801643653 \times 10^{-5}$ & $1.09426950315531$ & $1.80885790363103 \times 10^{-18}$ \\   
				
				0.5 & $10^{50}$ & $5.19393183842041 \times 10^{32}$ & $3.53187243391340 \times 10^{-18}$ & $1.09426952284475$ & $2.88087074816100 \times 10^{-95}$ \\     
				
				0.5 & $10^{100}$ & $5.19386368874490 \times 10^{66}$ & $3.53182661265498 \times 10^{-34}$ & $1.09426952284475$ & $2.88688588361371 \times 10^{-191}$ \\   
				
				\hline
		\end{tabular}}
\end{table}

\subsection{\textbf{C variation}}

Now, we want to investigate how the solutions to the system of Eqs. (\ref{T ponto R}), (\ref{evoaR}) and (\ref{evoBetaR}) depend on the
parameter $C$. Then, varying it and keeping all other parameters and initial conditions fixed, it was found that $a_{nc}$ expands more rapidly when we increase the value of $C$. In Figure 3, we present an example of that effect, in a graph of $a_{nc} \times t$, for four different values of $C$ ($C$ = 1, 10, 100 and 1000). Next, studying how the variation of $C$ influences the dynamics of $\beta_\pm$, we notice that for all values of $C$, $\beta_\pm$ always goes to a constant value after a period of expansion. The greater the value of $C$ the more rapidly $\beta_\pm$ goes to a constant value. In Figure 4, we present an example of that effect, in a graph of $\beta_{+} \times t$ (as explained, above, $\beta_{-}$ behaves in the same way), where we use four values of $C$ ($C$ = 1, 5, 10 and 100). In Table \ref{T2}, we present the values of the scale factors and their time derivatives for different times and $C$ values given in Figure 4. It is noted that the values of $\beta_{+} (t)$ stabilize while $\dot{\beta}_{+} (t)$ tend to zero, which characterizes the phenomenon of isotropization. On the other hand, $a_{nc} (t)$ is always expansive, although in a decelerated way, as shown by the values of $\dot{a}_{nc} (t)$.

\begin{figure}[!tbp]
	\begin{minipage}{0.4\textwidth}
		\centering
		\includegraphics[width=\linewidth]{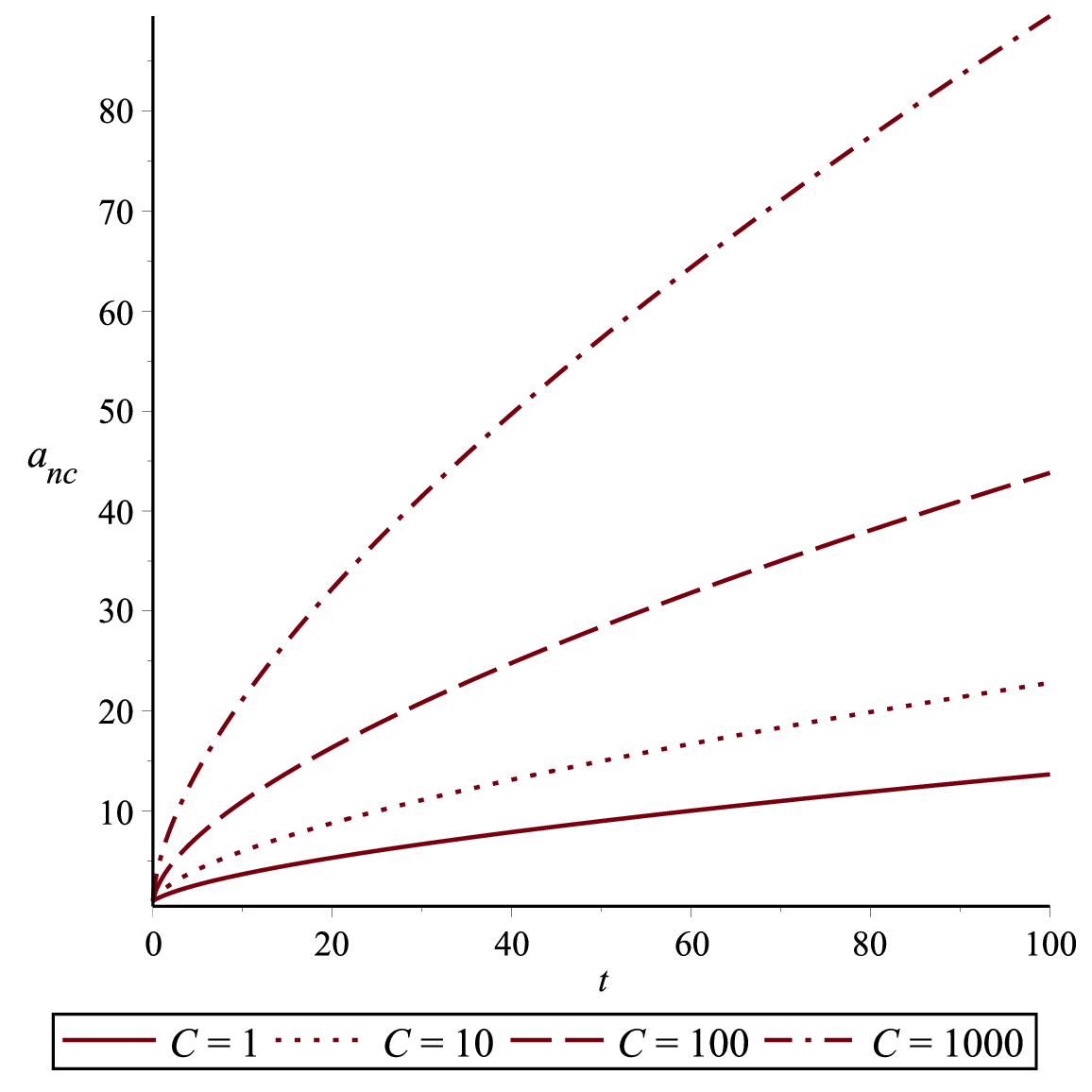}
		{\small Figure 3: Behavior of $a_{nc} (t)$ for different values of $C$. Here, we considered $C_+$ = $C_-$ = 0.1, $\dot{\beta}_{\pm} (0) = 0.026352314$ and $\gamma$ = $- 0.5$.}
		\label{AcomC}
	\end{minipage}\hfill
	\begin{minipage}{0.4\textwidth}
		\centering
		\includegraphics[width=\linewidth]{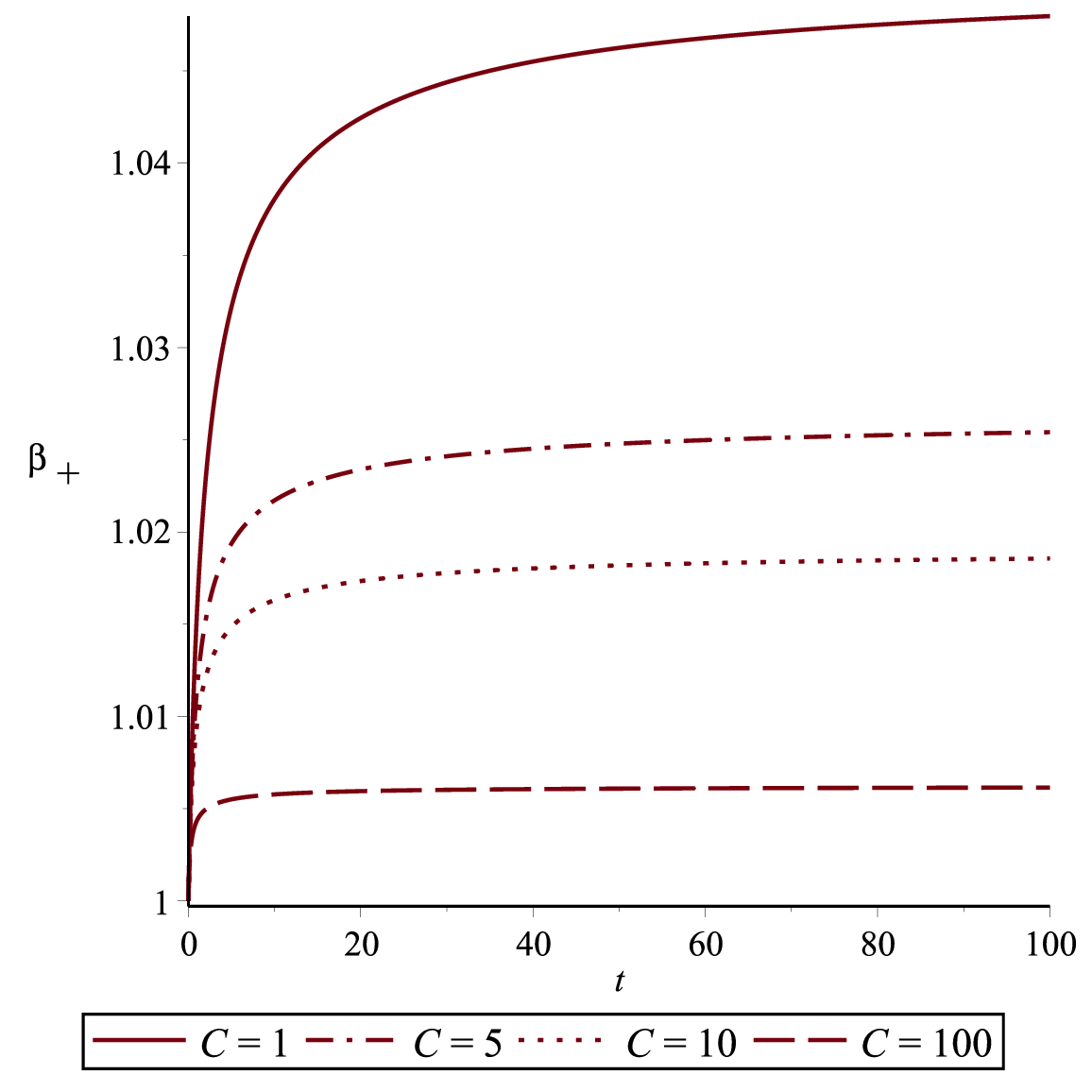}
		{\small Figure 4: Behavior of $\beta_{+} (t)$ for different values of $C$. Here, we considered $C_+$ = $C_-$ = 0.1, $\dot{\beta}_{\pm} (0) =  0.026352314$ and $\gamma$ = $- 0.5$.}
		\label{BcomC}
	\end{minipage}\hfill
\end{figure} 

\begin{table}[!tbp]
	\centering
	\caption{\small Values of $a_{nc}$, $\dot{a}_{nc}$, $\beta_{+}$ and $\dot{\beta}_{+}$ for different values of $C$ and $t$, with $C_+ = C_- = 0.1$, $\dot{\beta}_{\pm} (0) = 0.026352314$ and $\gamma = - 0.5$.}
	\label{T2}
		{\tiny\begin{tabular}{c c c c c c}
				\hline
				$C$ &  $t $ & $a_{nc} (t)$ & $\dot{a}_{nc} (t)$ & $\beta_{+} (t)$ & $\dot{\beta}_{+} (t)$ \\
				\hline

				1 & $10^{2}$ & $1.65125214932867 \times 10$  & $9.62641049235326 \times 10^{-2}$ & $1.04796809339111$ & $1.93548224946860 \times 10^{-5}$ \\
				
				1 & $10^{5}$ & $1.30963130232505 \times 10^{3}$ & $8.72306365248727 \times 10^{-3}$ & $1.05041676339555$ & $4.97086588375277 \times 10^{-11} $ \\
				
				1 & $10^{10}$ & $3.16302781606027 \times 10^{6}$ & $2.15011509337874 \times 10^{-4}$ & $1.05042217769972$ & $1.25483917402440 \times 10^{-20}$ \\
				
				1 & $10^{50}$ & $5.00973387704284 \times 10^{33}$ & $3.40661774744895 \times 10^{-17}$ & $1.05042217783612$ & $1.99208880401541 \times 10^{-97}$ \\
				
				1 & $10^{100}$ & $5.00966834808691 \times 10^{67}$ & $3.40657493410730 \times 10^{-33}$ & $1.05042217783612$ & $1.99629196108106 \times 10^{-193}$ \\ 
				
				\hline
				
				5 & $10^{2}$ & $2.07094584946924 \times 10$ & $1.22867097402338  \times 10^{-1}$ & $1.02540942871800$ & $6.97253148187132 \times 10^{-6}$ \\
				
				5 & $10^{5}$ & $1.77968245136721 \times 10^{3}$ & $1.19589992653161 \times 10^{-2}$ & $1.02628867731237$ & $1.78950880499895 \times 10^{-11}$ \\
				
				5 & $10^{10}$ & $4.37491221153109 \times 10^{6}$ & $2.97440204887313 \times 10^{-4}$ & $1.02629062675910$ & $4.51917528991155 \times 10^{-21}$ \\
				
				5 & $10^{50}$ & $6.93132780245766 \times 10^{33}$ & $4.71330124783900 \times 10^{-17}$ & $1.02629062680823$ & $7.17429471741857  \times 10^{-98}$ \\
				
				5 & $10^{100}$ & $6.93123722211913 \times 10^{67}$ & $ 4.71324206420892 \times 10^{-33}$ & $1.02629062680823$ & $7.18943966058331 \times 10^{-194}$ \\
				
				\hline
				
				10 & $10^{2}$ & $2.37730460483869 \times 10$ & $1.42585618763042 \times 10^{-1}$ & $1.01855834157711$ & $4.15189199923247 \times 10^{-6}$ \\
				
				10 & $10^{5}$ & $2.11463488755257 \times 10^{3}$ & $1.42531174013815 \times 10^{-2}$ & $1.01907781152268$ & $1.04448394588694 \times 10^{-11}$ \\
				
				10 & $10^{10}$ & $5.22938610117169 \times 10^{6}$ & $3.55553076621969 \times 10^{-4}$ & $1.01907894924070$ & $2.63703812222604 \times 10^{-21}$ \\
				
				10 & $10^{50}$ & $8.28594946527034 \times 10^{33}$ & $5.63444589906848 \times 10^{-17}$ & $1.01907894926937$ & $4.18648199541684 \times 10^{-98}$ \\
				
				10 & $10^{100}$ & $8.28584121325477 \times 10^{67}$ & $5.63436965504191 \times 10^{-33}$ & $1.01907894926937$ & $4.19512113247034 \times 10^{-194}$ \\
				
				\hline
				
				100 & $10^{2}$ & $4.25314375804034 \times 10$ & $2.65779980344301 \times 10^{-1}$ & $1.00614213548638$ & $6.27727247012018 \times 10^{-7}$ \\
				
				100 & $10^{5}$ & $4.15246296215656 \times 10^{3}$ & $2.81527793285521 \times 10^{-2}$ & $1.00621749072138$ & $ 1.41142196941052 \times 10^{-12}$ \\
				
				100 & $10^{10}$ & $1.03817419948622 \times 10^{7}$ & $7.05935752830522 \times 10^{-4}$ & $1.00621764434458$ & $3.55714087885152 \times 10^{-22}$ \\
				
				100 & $10^{50}$ & $1.64527786574461 \times 10^{34}$ & $1.11878889733860 \times 10^{-16}$ & $1.00621764434844$ & $5.64703066397212 \times 10^{-99}$ \\
				
				100 & $10^{100}$ & $1.64525653500275 \times 10^{68}$ & $1.11877463209467 \times 10^{-32}$ & $1.00621764434844$ & $5.65891800415214 \times 10^{-195}$ \\
				
				\hline
		\end{tabular}}
\end{table}

\subsection{\textbf{$C_\pm$ variation}}

Now, we want to investigate how the solutions to the system of Eqs. (\ref{T ponto R}), (\ref{evoaR}) and (\ref{evoBetaR}) depend on the
parameters $C_\pm$. Then, varying it and keeping all other parameters and initial conditions fixed, it was found that $a_{nc}$ expands more rapidly when we increase the values of $C_\pm$. In Figure 5, we present an example of that effect, in a graph of $a_{nc} \times t$, for four different values of $C_\pm$ ($C_\pm$ = 10, 100, 500 and 1000). Next, studying how the variations of $C_\pm$ influences the dynamics of $\beta_\pm$, we notice that for all values of $C_\pm$, $\beta_\pm$ always goes to a constant value after a period of expansion. It does not seem to have a clear pattern on how the values of $C_\pm$ influence how fast $\beta_\pm$ goes to a constant value.
In Figure 6, we present an example of that effect, in a graph of $\beta_{+} \times t$ (as explained, above, $\beta_{-}$ behaves in the same way), where we use five values of $C_\pm$ ($C_\pm$ = 1, 10, 25, 100 and 1000). In Table \ref{T3}, we present the values of the scale factors and their time derivatives for different times and the $C_\pm$ values given in Figure 6. It is noted that the values of $\beta_{+} (t)$ stabilize while $\dot{\beta}_{+} (t)$ tend to zero, which characterizes the phenomenon of isotropization. On the other hand, $a_{nc} (t)$ is always expansive, although in a decelerated way, as shown by the values of $\dot{a}_{nc} (t)$.

\begin{figure}[!tbp]
	\begin{minipage}{0.4\textwidth}
		\centering
		\includegraphics[width=\linewidth]{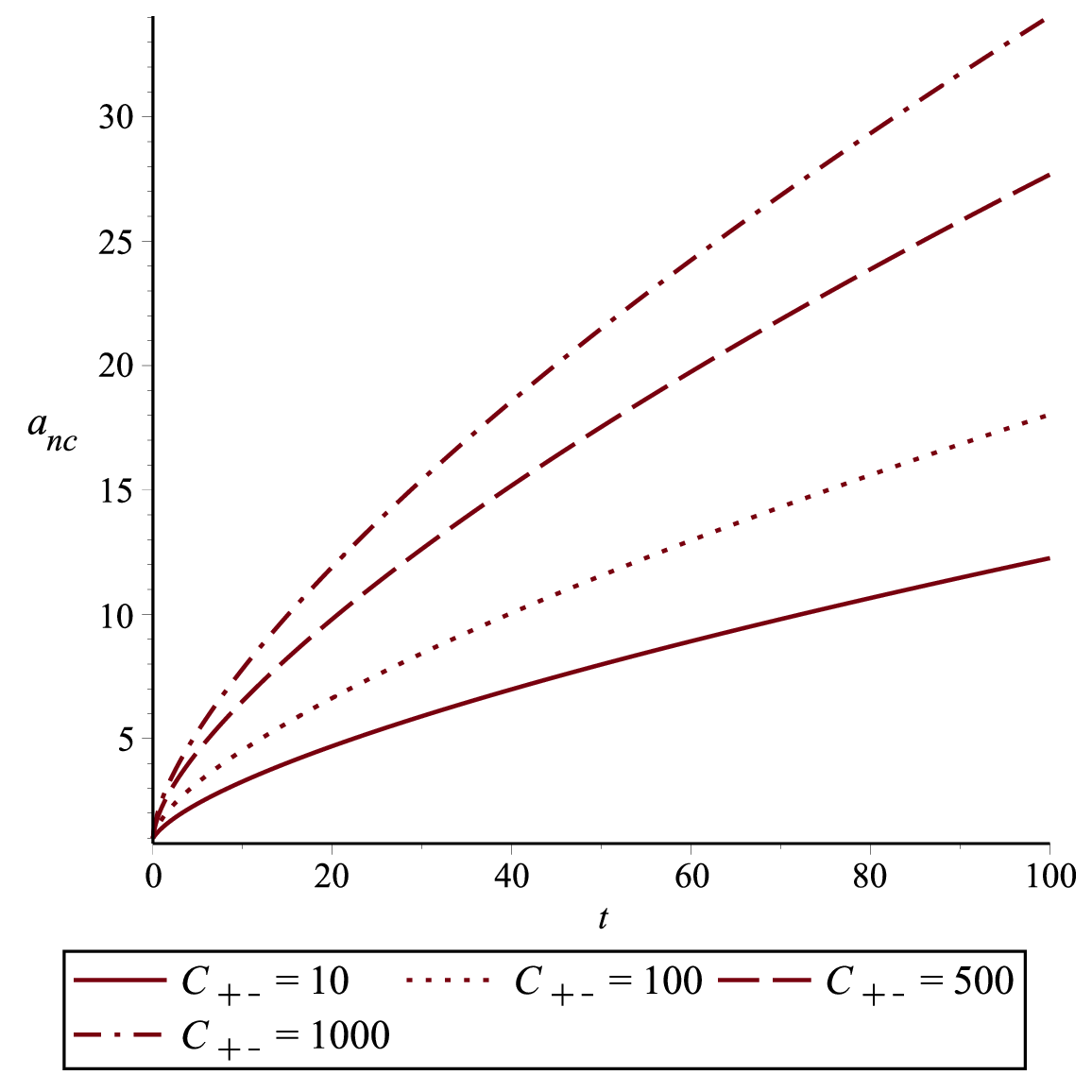}
		{\small Figure 5: Behavior of $a_{nc} (t)$ for different values of $C_\pm$. Here, we considered $C$ = 0.1, $C_+$ = $C_-$, $\dot{\beta}_{+} (0) = \dot{\beta}_{-} (0)$ and $\gamma$ = $- 0.5$.}
		\label{AcomC1}
	\end{minipage}\hfill
	\begin{minipage}{0.4\textwidth}
		\centering
		\includegraphics[width=\linewidth]{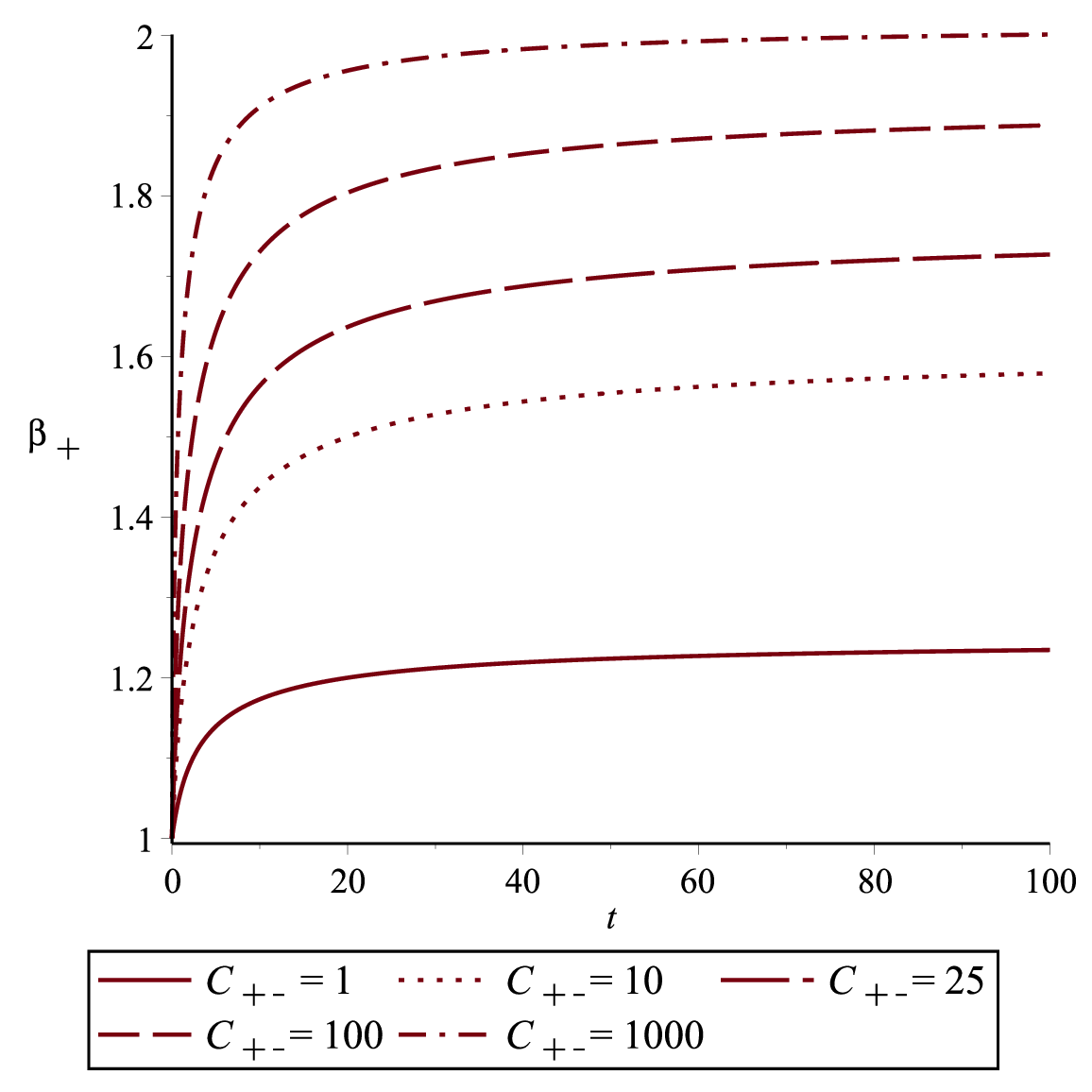}
		{\small Figure 6: Behavior of $\beta_{+} (t)$ for different values of $C_\pm$. Here, we considered $C$ = 0.1, $C_+$ = $C_-$, $\dot{\beta}_{+} (0) = \dot{\beta}_{-} (0)$ and $\gamma$ = $- 0.5$.}
		\label{BcomC1}
	\end{minipage}\hfill
\end{figure} 

\begin{table}[!tbp]
		\centering
		\caption{\small Values of $a_{nc}$, $\dot{a}_{nc}$, $\beta_+$ = $\beta_-$ and $\dot{\beta}_{+}$ = $\dot{\beta}_{-}$ for different values of $C_\pm$ and $t$, with $C = 0.1$, $\dot{\beta}_{+} (0) = \dot{\beta}_{-} (0)$ and $\gamma = - 0.5$.}
		\label{T3}
		{\tiny\begin{tabular}{c c c c c c}
				\hline
				$C_{\pm}$ &  $t $ & $a_{nc} (t)$ & $\dot{a}_{nc} (t)$ & $\beta_{+} (t)$ & $\dot{\beta}_{+} (t)$  \\
				\hline

				1 & $10^{2}$ & $1.48163577331849 \times 10$ & $8.59541006226603 \times 10^{-2}$ & $1.23429180777149$ & $1.21109858942679\times 10^{-4}$ \\
				
				1 & $10^{5}$ & $1.10953012701424 \times 10^{3}$ & $7.33221507214260 \times 10^{-3}$ & $1.24934366871540$ & $2.92647847309175 \times 10^{-10}$ \\
				
				1 & $10^{10}$ & $2.63725110493034 \times 10^{6}$ & $1.79243890946981 \times 10^{-4}$ & $1.24937552377710$ & $7.37657808961067 \times 10^{-20}$ \\
				
				1 & $10^{50}$ & $4.17578841748877 \times 10^{33}$ & $2.83953665623043 \times 10^{-17}$ & $1.24937552457892$ & $1.17106511180275 \times 10^{-96}$ \\
				
				1 & $10^{100}$ & $4.17573420025791 \times 10^{67}$ & $2.83949883781988 \times 10^{-33}$ & $1.24937552457892$ & $1.17349602320445 \times 10^{-192}$ \\

				\hline

				10 & $10^{2}$ & $1.55932186538612 \times 10$ & $9.21250131792895 \times 10^{-2}$ & $1.57887093271609$ & $2.70518842710661 \times 10^{-4}$ \\
				
				10 & $10^{5}$ & $1.24719915969613 \times 10^{3}$ & $8.29414045474369 \times 10^{-3}$ & $1.61156140429005$ & $6.03343606261464 \times 10^{-10}$ \\
				
				10 & $10^{10}$ & $3.00125948852418 \times 10^{6}$ & $2.04007238282992 \times 10^{-4}$ & $1.61162704235069$ & $1.51882716022119 \times 10^{-19}$ \\
				
				10 & $10^{50}$ & $4.75317464071994 \times 10^{33}$ & $3.23215739578492 \times 10^{-17}$ & $1.61162704400166$ & $2.41119018220561 \times 10^{-96}$ \\
				
				10 & $10^{100}$ & $4.75311232276144 \times 10^{67}$ & $3.23211641634281 \times 10^{-33}$ & $1.61162704400166$ & $2.41626092390146 \times 10^{-192}$ \\
				
				\hline

				25 & $10^{2}$ & $1.66070967568793 \times 10$ & $9.98208161301555 \times 10^{-2}$ & $1.72690823482933$ & $3.00149908008377 \times 10^{-4}$ \\
				
				25 & $10^{5}$ & $1.40431187211007 \times 10^{3}$ & $9.38495703035183 \times 10^{-3}$ & $1.76239290543430$ & $6.31370450433692 \times 10^{-10}$ \\
				
				25 & $10^{10}$ & $3.41175848264912 \times 10^{6}$ & $2.31930650844247 \times 10^{-4}$ & $1.76246157000559$ & $1.58818808870899 \times 10^{-19}$ \\
				
				25 & $10^{50}$ & $5.40418075145210 \times 10^{33}$ & $3.67484160650540 \times 10^{-17}$ & $1.76246157173194$ & $2.52128283221340 \times 10^{-96}$ \\
				
				25 & $10^{100}$ & $5.40411012067885 \times 10^{67}$ & $3.67479545987391 \times 10^{-33}$ & $1.76246157173194$ & $2.52660501893193 \times 10^{-192}$ \\
				
				\hline
				
				100 & $10^{2}$ & $1.98922119578162 \times 10$ & $1.23814629473104 \times 10^{-1}$ & $1.887935924385113$ & $2.64030880422531 \times 10^{-4}$ \\
				
				100 & $10^{5}$ & $1.85034089921554 \times 10^{3}$ & $1.24591486795781 \times 10^{-2}$ & $1.91812059889195$ & $5.11599716231378 \times 10^{-10}$ \\
				
				100 & $10^{10}$ & $4.56124988782102 \times 10^{6}$ & $3.10113451865018 \times 10^{-4}$ & $1.91817621820345$ & $1.28588446091742 \times 10^{-19}$ \\
				
				100 & $10^{50}$ & $ 7.22674859798064 \times 10^{33}$ & $4.91418735789468 \times 10^{-17}$ & $1.91817621960119$ & $2.04136776328131 \times 10^{-96}$ \\
				
				100 & $10^{100}$ & $7.22665418185123 \times 10^{67}$ & $4.91412565783232 \times 10^{-33}$ & $1.91817621960119$ & $2.04567751961172 \times 10^{-192}$ \\
				
				\hline
				
				1000 & $10^{2}$ & $3.37582307751617 \times 10$ & $2.21267441438632 \times 10^{-1}$ & $2.00101413700075$ & $1.300386911706965 \times 10^{-4}$ \\
				
				1000 & $10^{5}$ & $3.49996371309542 \times 10^{3}$ & $2.37347030068579 \times 10^{-2}$ & $2.01539286212008$ & $2.344859657581494 \times 10^{-10}$ \\	 		
				
				1000 & $10^{10}$ & $8.74607614611653 \times 10^{6}$ & $5.94706195229799 \times 10^{-4}$ & $2.01541835078982$ & $5.89161382382453 \times 10^{-20}$ \\
				
				1000 & $10^{50}$ & $1.38602689533257 \times 10^{34}$ & $9.42497896358757 \times 10^{-17}$ & $2.01541835143024$ & $9.35311256235490 \times 10^{-97}$ \\
				
				1000 & $10^{100}$ & $1.38600873357635 \times 10^{68}$ & $9.42485989442013 \times 10^{-33}$ & $2.01541835143024$ & $9.37280647608427 \times 10^{-193}$ \\
				
				\hline
		\end{tabular}}
	\end{table}

	\subsection{\textbf{$a_c (0)$ variation}}
	
Now, we want to investigate how the solutions to the system of Eqs. (\ref{T ponto R}), (\ref{evoaR}) and (\ref{evoBetaR}) depend on the initial
condition $a_c (0)$. Then, varying it and keeping all other parameters and initial conditions fixed, it was found that $a_{nc}$ expands more rapidly for smaller values of $a_c (0)$. In Figure 7, we present an example of that effect, in a graph of $a_{nc} \times t$, for four different values of $a_c (0)$ ($a_c (0)$ = 0.5, 1, 5 and 10). Next, studying how the variation of $a_c (0)$ influences the dynamics of $\beta_\pm$, we notice that for all values of $a_c (0)$, $\beta_\pm$ always goes to a constant value after a period of expansion. The smaller the value of $a_c (0)$ the more rapidly $\beta_\pm$ goes to a constant value. In Figure 8, we present an example of that effect, in a graph of $\beta_{+} \times t$ (as explained, above, $\beta_{-}$ behaves in the same way), where we use five different values of $a_c (0)$ ($a_c (0)$ = 0.5, 0.75, 1, 2.5 and 5). In Table \ref{T4}, we present the values of the scale factors and their time derivatives for different times and $a_c (0)$ values given in Figure 8. It is noted that the values of $\beta_{+} (t)$ stabilize while $\dot{\beta}_{+} (t)$ tend to zero, which characterizes the phenomenon of isotropization. On the other hand, $a_{nc} (t)$ is always expansive, although in a decelerated way, as shown by the values of $\dot{a}_{nc} (t)$.

\begin{figure}[!tbp]
	\begin{minipage}{0.4\textwidth}
		\centering
		\includegraphics[width=\linewidth]{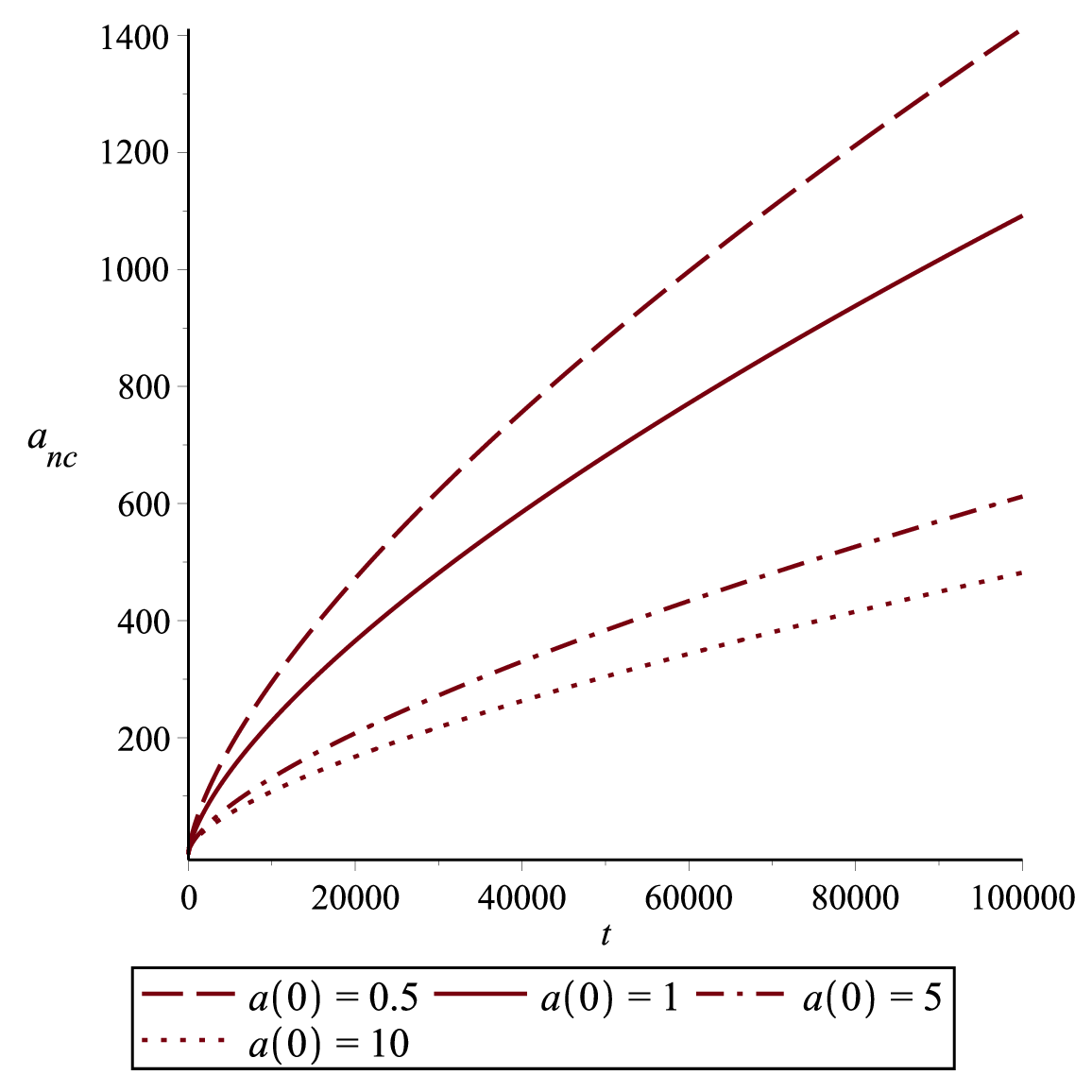}
		{\small Figure 7: Behavior of $a_{nc} (t)$ for different values of $a_{c} (0)$. Here, we considered $C$ = $C_+$ = $C_-$ = 0.1, $\dot{\beta}_{+} (0) = \dot{\beta}_{-} (0)$ and $\gamma$ = $- 0.5$.}
		\label{AcomA0}
	\end{minipage}\hfill
	\begin{minipage}{0.4\textwidth}
		\centering
		\includegraphics[width=\linewidth]{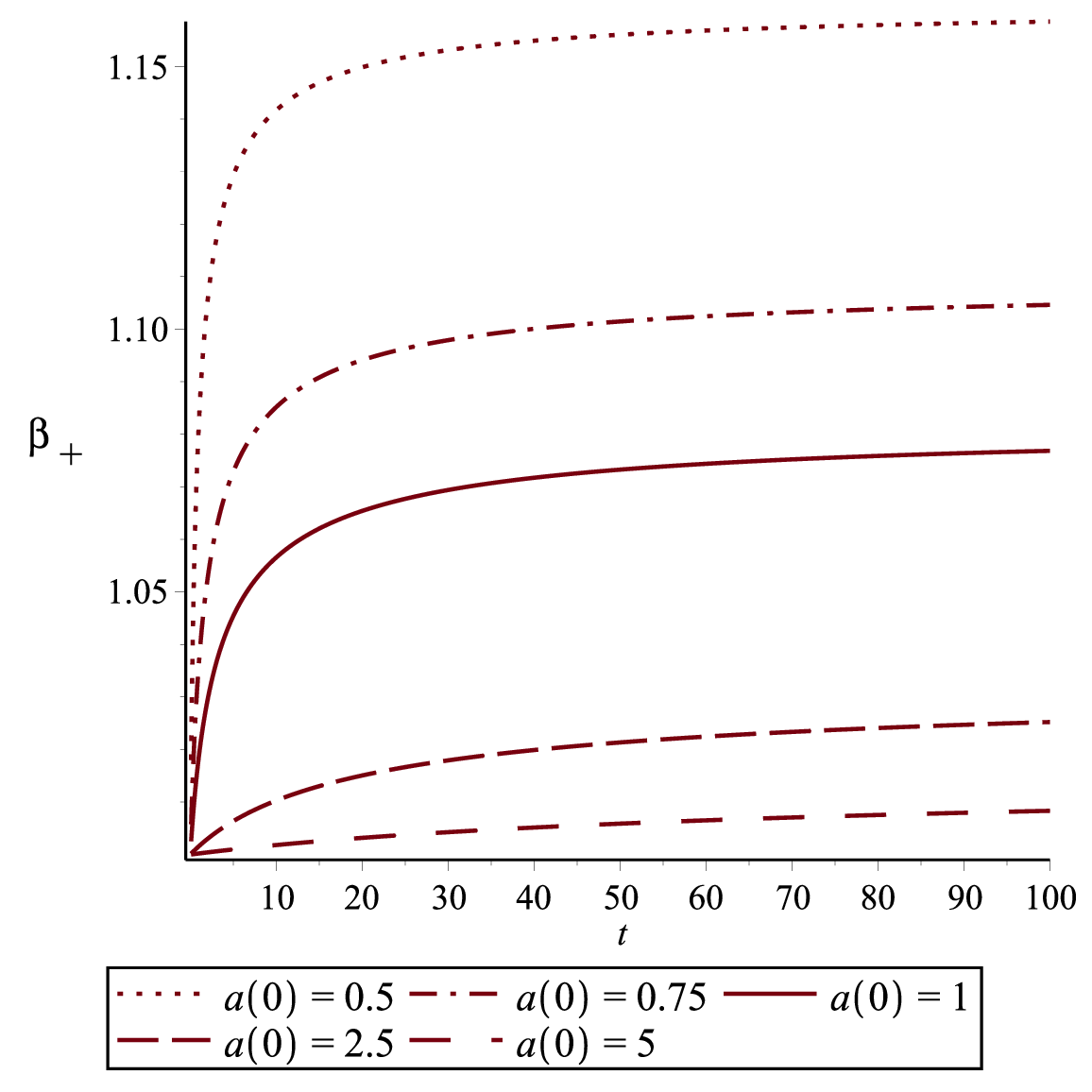}
		{\small Figure 8: Behavior of $\beta_{+} (t)$ for different values of $a_{c} (0)$. Here, we considered $C$ = $C_+$ = $C_-$ = 0.1, $\dot{\beta}_{+} (0) = \dot{\beta}_{-} (0)$ and $\gamma$ = $- 0.5$.}
		\label{BcomA0}
	\end{minipage}\hfill
\end{figure} 

\begin{table}[!tbp]
	\centering
	\caption{\small Values of $a_{nc}$, $\dot{a}_{nc}$, $\beta_{+}$ and $\dot{\beta}_{+}$ for different values of $a_{c} (0)$ and $t$, with $C = C_+ = C_- = 0.1$, $\dot{\beta}_{+} (0) = \dot{\beta}_{-} (0)$ and $\gamma = - 0.5$.}
	\label{T4}
		{\tiny\begin{tabular}{c c c c c c}
				\hline
				$a_{c} (0)$ &  $t$ & $a_{nc} (t)$ & $\dot{a}_{nc} (t)$ & $\beta_{+} (t)$ & $\dot{\beta}_{+} (t)$  \\
				\hline
				
				0.5 & $10^{2}$ & $1.66671229768871 \times 10$ & $9.95112019035336 \times 10^{-2}$ & $1.15854230412839$ & $2.67954956286370 \times 10^{-5}$ \\
				
				0.5 & $10^{5}$ & $1.37847054891445 \times 10^{3}$ & $9.20258194728948 \times 10^{-3}$ & $1.16163472887432$ & $5.39392401410827 \times 10^{-11}$ \\							
				
				0.5 & $10^{10}$ & $3.34322590764208 \times 10^{6}$ & $2.27269035988505 \times 10^{-4}$ & $1.16164059509821$ & $1.35689601850210 \times 10^{-20}$ \\							
				
				0.5 & $10^{50}$ & $5.29550140634861 \times 10^{33}$ & $3.60094134704791 \times 10^{-17}$ & $1.16164059524570$ & $2.15410916395753 \times 10^{-97}$ \\ 							
				
				0.5 & $10^{100}$ & $5.29543289969475 \times 10^{67}$ & $3.60089476126396 \times 10^{-33}$ & $1.16164059524570$ & $2.15862442483626 \times 10^{-193}$ \\							
				
				\hline
				
				0.75 & $10^{2}$ & $1.54475101988563 \times 10$ & $9.04778642991938 \times 10^{-2}$ & $1.10462092982711$ & $3.46798280409981 \times 10^{-5}$ \\
				
				0.75 & $10^{5}$ & $1.20006188402228 \times 10^{3}$ & $7.96369839149115 \times 10^{-3}$ & $1.10878361513688$ & $7.69385477805382 \times 10^{-11}$ \\							
				
				0.75 & $10^{10}$ & $2.87645291091118 \times 10^{6}$ & $1.95516965487555 \times 10^{-4}$ & $1.10879198646910$ & $1.93743214273057 \times 10^{-20}$ \\  							
				
				0.75 & $10^{50}$ & $4.55522111633149 \times 10^{33}$ & $3.09754905605154 \times 10^{-17}$ & $1.10879198667970$ & $3.07574158389866 \times 10^{-97}$ \\ 	   							
				
				0.75 & $10^{100}$ & $4.55516138833136 \times 10^{67}$ & $3.09750975361060 \times 10^{-33}$ & $1.10879198667970$ & $3.08220824607727 \times 10^{-193}$ \\ 							
				
				\hline

				1 & $10^{2}$ & $1.47261847316966 \times 10$ & $8.52059345807472 \times 10^{-2}$ & $1.07684443967982$ & $4.01814179551821 \times 10^{-5}$ \\
				
				1 & $10^{5}$ & $1.09169175591580 \times 10^{3}$ & $7.20701293489404 \times 10^{-3}$ & $1.08186070638515$ & $9.83964209788674 \times 10^{-11}$ \\	 							
				
				1 & $10^{10}$ & $ 2.58968680211751 \times 10^{6}$ & $1.76007813141873 \times 10^{-4}$ & $1.08187141805590$ & $2.48074434412398 \times 10^{-20}$ \\  	 							
				
				1 & $10^{50}$ & $4.10033296508804 \times 10^{33}$ & $2.78822525402685 \times 10^{-17}$ & $1.08187141832555$ & $3.93827793662011 \times 10^{-97}$ \\   							
				
				1 & $10^{100}$ & $4.10027918315254 \times 10^{67}$ & $2.78818974557760 \times 10^{-33}$ & $1.08187141832555$ & $3.94654805491996 \times 10^{-193}$ \\    							
				
				\hline
				
				2.5 & $10^{2}$ & $1.30205078741326 \times 10$ & $7.17905722395018 \times 10^{-2}$ & $1.02522093590183$ & $4.86798375085851 \times 10^{-5}$ \\
				
				2.5 & $10^{5}$ & $8.23256178087734 \times 10^{2}$ & $5.31003058430142 \times 10^{-3}$ & $1.03284688661337$ & $2.11728352840473 \times 10^{-10}$ \\ 							
				
				2.5 & $10^{10}$ & $1.86041992921103 \times 10^{6}$ & $1.26382136988052 \times 10^{-4}$ & $1.03287002986384$ & $5.38861036137084 \times 10^{-20}$ \\    							
				
				2.5 & $10^{50}$ & $2.94296372779180 \times 10^{33}$ & $2.00121546600519 \times 10^{-17}$ & $1.03287003044957$ & $8.55457584482055 \times 10^{-97}$ \\      							
				
				2.5 & $10^{100}$ & $2.94292564369923 \times 10^{67}$ & $2.00118968398000 \times 10^{-33}$ & $1.03287003044957$ & $8.57253095289013 \times 10^{-193}$ \\    							
				
				\hline
				
				5 & $10^{2}$ & $1.26404758754218 \times 10$ & $6.00684593176709 \times 10^{-2}$ & $1.00832426316329$ & $3.55400966748681 \times 10^{-5}$ \\
				
				5 & $10^{5}$ & $6.80509716581963 \times 10^{2}$ & $4.27885614016897 \times 10^{-3}$ & $1.01640278779626$ & $3.71552594819910 \times 10^{-10}$ \\ 							
				
				5 & $10^{10}$ &	$1.45097371432128 \times 10^{6}$ & $9.85040796527953 \times 10^{-5}$ & $1.01644376066946$ & $9.65379063007719 \times 10^{-20}$ \\     						
				
				5 & $10^{50}$ &	$2.29246567379484 \times 10^{33}$ & $1.55887604531451 \times 10^{-17}$ & $1.01644376171883$ & $1.53257868073372 \times 10^{-96}$ \\   						
				
				5 & $10^{100}$ & $2.29243567034892 \times 10^{67}$ & $1.55885643309785 \times 10^{-33}$ & $1.01644376171883$ & $1.53581077320607 \times 10^{-192}$ \\  											
				
				\hline
		\end{tabular}}
\end{table}

\subsection{\textbf{$\dot{a}_{c} (0)$ variation}}

Now, we want to investigate how the solutions to the system of Eqs. (\ref{T ponto R}), (\ref{evoaR}) and (\ref{evoBetaR}) depend on the initial
condition $\dot{a}_{c} (0)$. Then, varying it and keeping all other parameters and initial conditions fixed, it was found that $a_{nc}$ expands more rapidly for greater values of $\dot{a}_{c} (0)$. In Figure 9, we present an example of that effect, in a graph of $a_{nc} \times t$, for five different values of $\dot{a}_{c} (0)$ ($\dot{a}_{c} (0)$ = 1, 2.5, 5, 7.5 and 10). Next, studying how the variation of $\dot{a}_{c} (0)$ influences the dynamics of $\beta_\pm$, we notice that for all values of $\dot{a}_{c} (0)$, $\beta_\pm$ always goes to a constant value after a period of expansion. The greater the value of $\dot{a}_{c} (0)$ the more rapidly $\beta_\pm$ goes to a constant value. In Figure 10, we present an example of that effect, in a graph of $\beta_{+} \times t$ (as explained, above, $\beta_{-}$ behaves in the same way), where we use four different values of $\dot{a}_{c} (0)$ ($\dot{a}_{c} (0)$ = 1, 2.5, 5 and 10). In Table \ref{T5}, we present the values of the scale factors and their time derivatives for different times and $\dot{a}_{c} (0)$ values given in Figure 10. It is noted that the values of $\beta_{+} (t)$ stabilize while $\dot{\beta}_{+} (t)$ tend to zero, which characterizes the phenomenon of isotropization. On the other hand, $a_{nc} (t)$ is always expansive, although in a decelerated way, as shown by the values of $\dot{a}_{nc} (t)$.

\begin{figure}[!tbp]
	\begin{minipage}{0.4\textwidth}
		\centering
		\includegraphics[width=\linewidth]{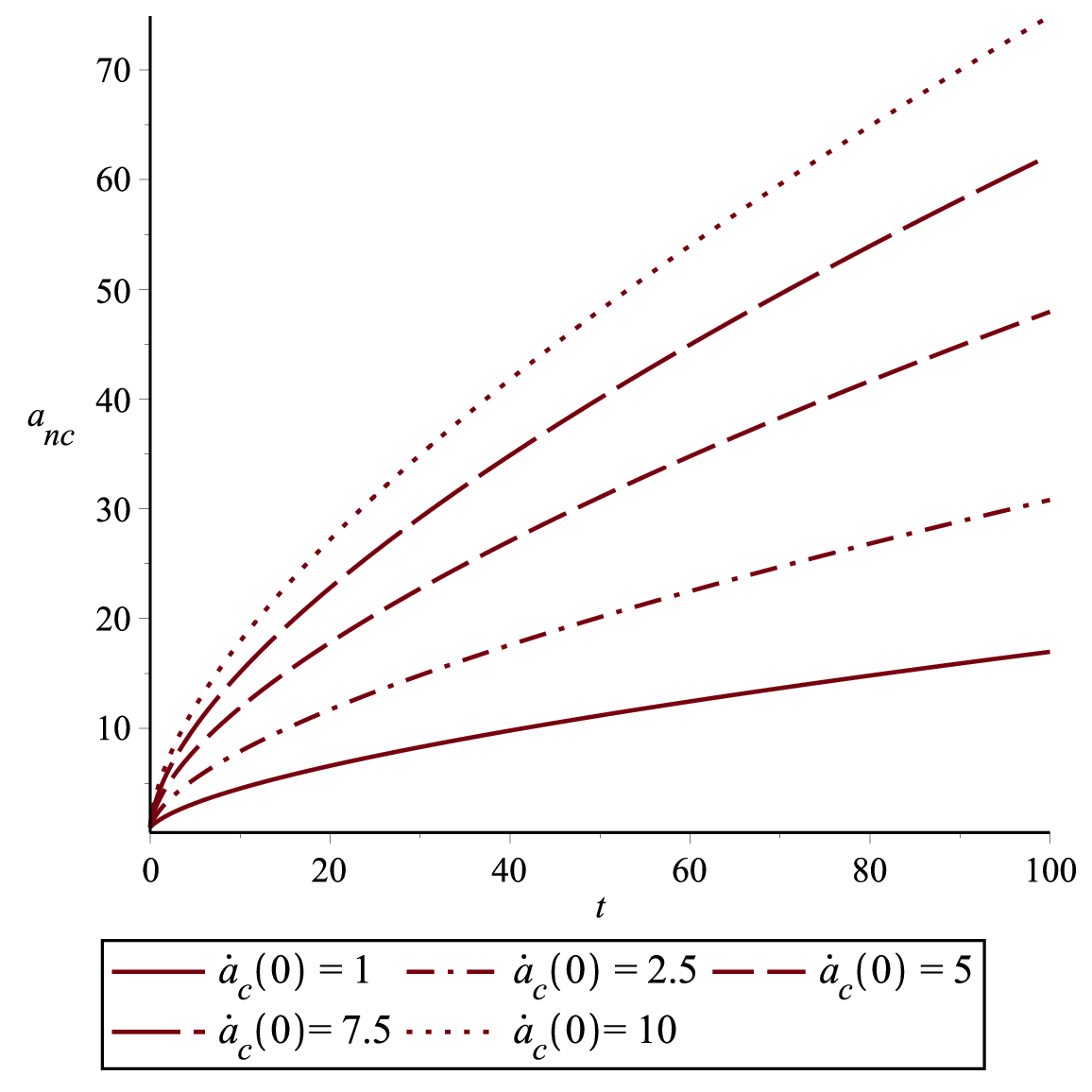}
		{\small Figure 9: Behavior of $a_{nc} (t)$ for different values of $\dot{a}_{c} (0)$. Here, we considered $C$ = $C_+$ = $C_-$ = 0.1, $\dot{\beta}_{\pm} (0) = 0.026352314$ and $\gamma$ = $- 0.5$.}
		\label{Acomv}
	\end{minipage}\hfill
	\begin{minipage}{0.4\textwidth}
		\centering
		\includegraphics[width=\linewidth]{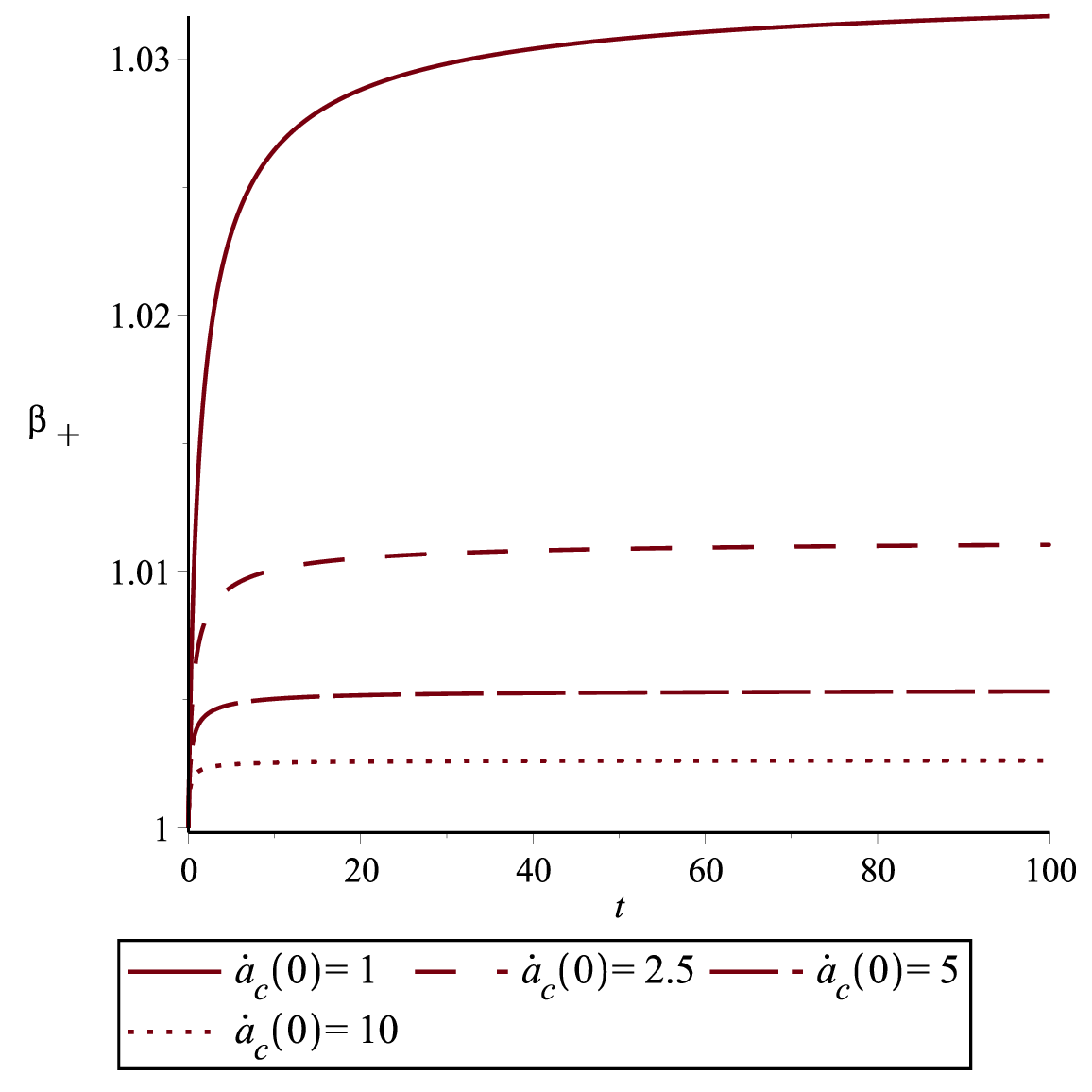}
		{\small Figure 10: Behavior of $\beta_{+} (t)$ for different values of $\dot{a}_{c} (0)$. Here, we considered $C$ = $C_+$ = $C_-$ = 0.1, $\dot{\beta}_{\pm} (0) = 0.026352314$ and $\gamma$ = $- 0.5$.}
		\label{Bcomv}
	\end{minipage}\hfill
\end{figure}

\begin{table}[!tbp]
	\centering
	\caption{\small Values of $a_{nc}$, $\dot{a}_{nc}$, $\beta_{+}$ and $\dot{\beta}_{+}$ for different values of $\dot {a}_{c} (0)$ and $t$, with $\dot{\beta}_{\pm} (0) = 0.026352314$ and $\gamma$ = - 0.5.}
	\label{T5}
		{\tiny\begin{tabular}{c c c c c c c c}
				\hline
				$\dot{a}_{c} (0)$ &  $t$ & $a_{nc} (t)$ & $\dot{a}_{nc} (t)$ & $\beta_{+} (t)$ & $\dot{\beta}_{+} (t)$  \\
				\hline
				
				1 & $10^{2}$ & $1.89779179633412 \times 10$ & $1.11818805826265 \times 10^{-1}$ & $1.03169011206471$ & $9.98605075778449 \times 10^{-6}$ \\
				
				1 & $10^{5}$ & $1.58846853984890 \times 10^{3}$ & $1.06459250820241 \times 10^{-2}$ & $1.03295362638406$ & $2.58168325987133 \times 10^{-11}$ \\
				
				1 & $10^{10}$ & $3.88442660381063 \times 10^{6}$ & $2.64080317544402\times 10^{-4}$ & $1.03295643885024$ & $6.51982013124600 \times 10^{-21}$ \\
				
				1 & $10^{50}$ & $6.15366986324990 \times 10^{33}$ & $4.18449501197903 \times 10^{-17}$ & $1.03295643892111$ & $1.03506550414104 \times 10^{-97}$ \\  
				
				1 & $10^{100}$ & $6.15358928186980 \times 10^{67}$ & $4.18443912392307 \times 10^{-33}$ & $1.03295643892111$ & $1.03720873872852 \times 10^{-193}$ \\ 
				
				\hline
				
				2.5 & $10^{2}$ & $3.07665231928596 \times 10$ & $1.88143345494552\times 10^{-1}$ & $1.01103339807890$ & $1.72850100795333 \times 10^{-6}$ \\
				
				2.5 & $10^{5}$ & $2.87460687107742 \times 10^{3}$ & $1.94436892459576 \times 10^{-2}$ & $1.01124579269028$ & $4.14146386397607 \times 10^{-12}$ \\
				
				2.5 & $10^{10}$ & $7.15645514123451 \times 10^{6}$ & $4.86605831911691 \times 10^{-4}$ & $1.01124624364767$ & $1.04478032072023 \times 10^{-21}$ \\
				
				2.5 & $10^{50}$ & $1.13406483908137 \times 10^{34}$ & $7.71164166756229 \times 10^{-17}$ & $1.01124624365903$ & $1.65865756327414 \times 10^{-98}$ \\
				
				2.5 & $10^{100}$ & $1.13405004108351 \times 10^{68}$ & $7.71153717714383 \times 10^{-33}$ & $1.01124624365903$ & $1.66207992901653 \times 10^{-194}$ \\  
				
				\hline
				
				5 & $10^{2}$ & $4.63356259514685 \times 10$ & $2.91051581313243 \times 10^{-1}$ & $1.00530069571624$ & $4.84599188884677 \times 10^{-7}$ \\
				
				5 & $10^{5}$ & $4.56616837736691 \times 10^{3}$ & $3.09700593590186\times 10^{-2}$ & $1.00535852910392$ & $1.07248034052911 \times 10^{-12}$ \\
				
				5 & $10^{10}$ & $1.14240923643710 \times 10^{7}$ & $7.76817369262152 \times 10^{-4}$ & $1.00535864582320$ & $2.70221181564066 \times 10^{-22}$ \\
				
				5 & $10^{50}$ & $1.81048676198374 \times 10^{34}$ & $1.23113048805118 \times 10^{-16}$ & $1.00535864582614$ & $4.28984299091916 \times 10^{-99}$ \\
				
				5 & $10^{100}$ & $1.81046304181695 \times 10^{68}$ & $1.23111494579452 \times 10^{-32}$ & $1.00535864582614$ & $4.29887798597234 \times 10^{-195}$ \\   
				
				\hline
				
				10 & $10^{2}$ & $7.13058264263282 \times 10$ & $4.58050886977793 \times 10^{-1}$ & $1.00260175471771$ & $1.36054680081620 \times 10^{-7}$ \\
				
				10 & $10^{5}$ & $7.28719549058469 \times 10^{3}$ & $4.94888013451013\times 10^{-2}$ & $1.00261757066591$ & $2.80768639626193 \times 10^{-13}$ \\ 
				
				10 & $10^{10}$ & $1.82710345323575 \times 10^{7}$ & $1.24241714148766 \times 10^{-3}$ & $1.00261760120904$ & $7.06719774223175 \times 10^{-23}$ \\
				
				10 & $10^{50}$ & $2.89567642560805 \times 10^{34}$ & $1.96906020583791 \times 10^{-16}$ & $1.00261760120980$ & $1.12195722166744 \times 10^{-99}$ \\
				
				10 & $10^{100}$ & $2.89563869940331 \times 10^{68}$ & $1.96903365972069 \times 10^{-32}$ & $1.00261760120980$ & $1.12427634267368 \times 10^{-195}$ \\   
				
				\hline
		\end{tabular}}
\end{table}

\subsection{\textbf{$\dot{\beta}_{\pm} (0)$ variation}}

Now, we want to investigate how the solutions to the system of Eqs. (\ref{T ponto R}), (\ref{evoaR}) and (\ref{evoBetaR}) depend on the initial
condition $\dot{\beta}_{\pm} (0)$. Then, varying it and keeping all other parameters and initial conditions fixed, it was found that $a_{nc}$ expands more rapidly for greater values of $\dot{\beta}_{\pm} (0)$. In Figure 11, we present an example of that effect, in a graph of $a_{nc} \times t$, for five different values of $\dot{\beta}_{+} (0)$ ($\dot{\beta}_{+} (0)$ = 1, 2.5, 5, 7.5 and 10). Next, studying how the variation of $\dot{\beta}_{\pm} (0)$ influences the dynamics of $\beta_\pm$, we notice that for all values of $\dot{\beta}_{\pm} (0)$, $\beta_\pm$ always goes to a constant value after a period of expansion. We may add to that general result the following behaviors: (i) if $\dot{\beta}_{\pm} (0) < 1$, the smaller the value of $\dot{\beta}_{\pm} (0)$ the more rapidly $\beta_\pm$ goes to a constant value; and (ii) if $\dot{\beta}_{\pm} (0) \geq 1$, the greater the value of $\dot{\beta}_{\pm} (0)$ the more rapidly $\beta_\pm$ goes to a constant value. In Figure 12, we present an example of that effect, in a graph of $\beta_{+} \times t$ (as explained, above, $\beta_{-}$ behaves in the same way), where we use five different values of $\dot{\beta}_{+} (0)$ ($\dot{\beta}_{+} (0)$ = 0.1, 0.5, 1, 2.5 and 10). In Table \ref{T6}, we present the values of the scale factors and their time derivatives for different times and $\dot{\beta}_{+} (0)$ values given in Figure 12. It is noted that the values of $\beta_{+} (t)$ stabilize while $\dot{\beta}_{+} (t)$ tend to zero, which characterizes the phenomenon of isotropization. On the other hand, $a_{nc} (t)$ is always expansive, although in a decelerated way, as shown by the values of $\dot{a}_{nc} (t)$.

\begin{figure}[!tbp]
	\begin{minipage}{0.4\textwidth}
		\centering
		\includegraphics[width=\linewidth]{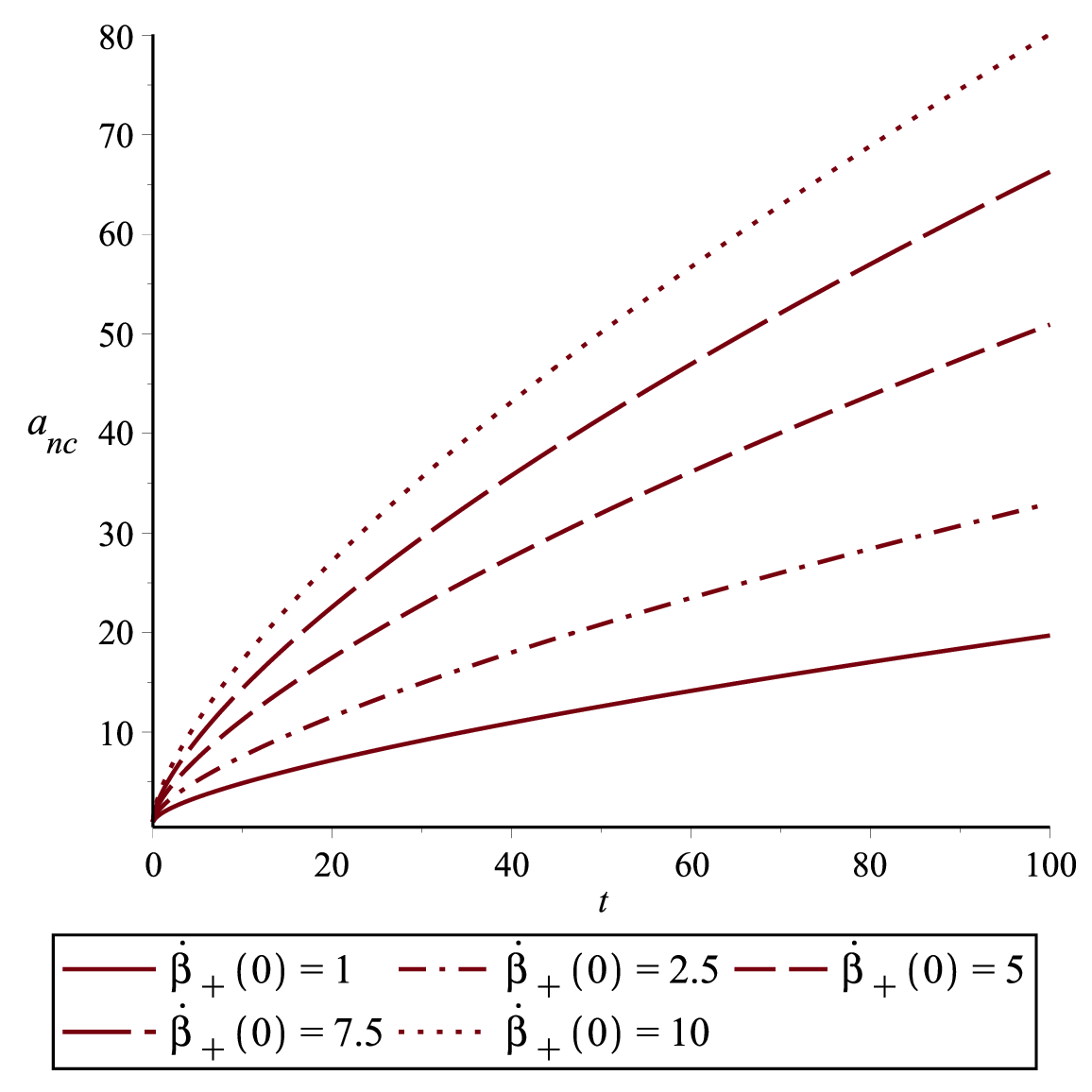}
		{\small Figure 11: Behavior of $a_{nc} (t)$ for different values of $\dot{\beta}_{+} (0)$ = $\dot{\beta}_{-} (0)$. Here, we considered $C$ = 0.1, $C_-$ = $C_+$, and $\gamma$ = $- 0.5$.}
		\label{Acomv1}
	\end{minipage}\hfill
	\begin{minipage}{0.4\textwidth}
		\centering
		\includegraphics[width=\linewidth]{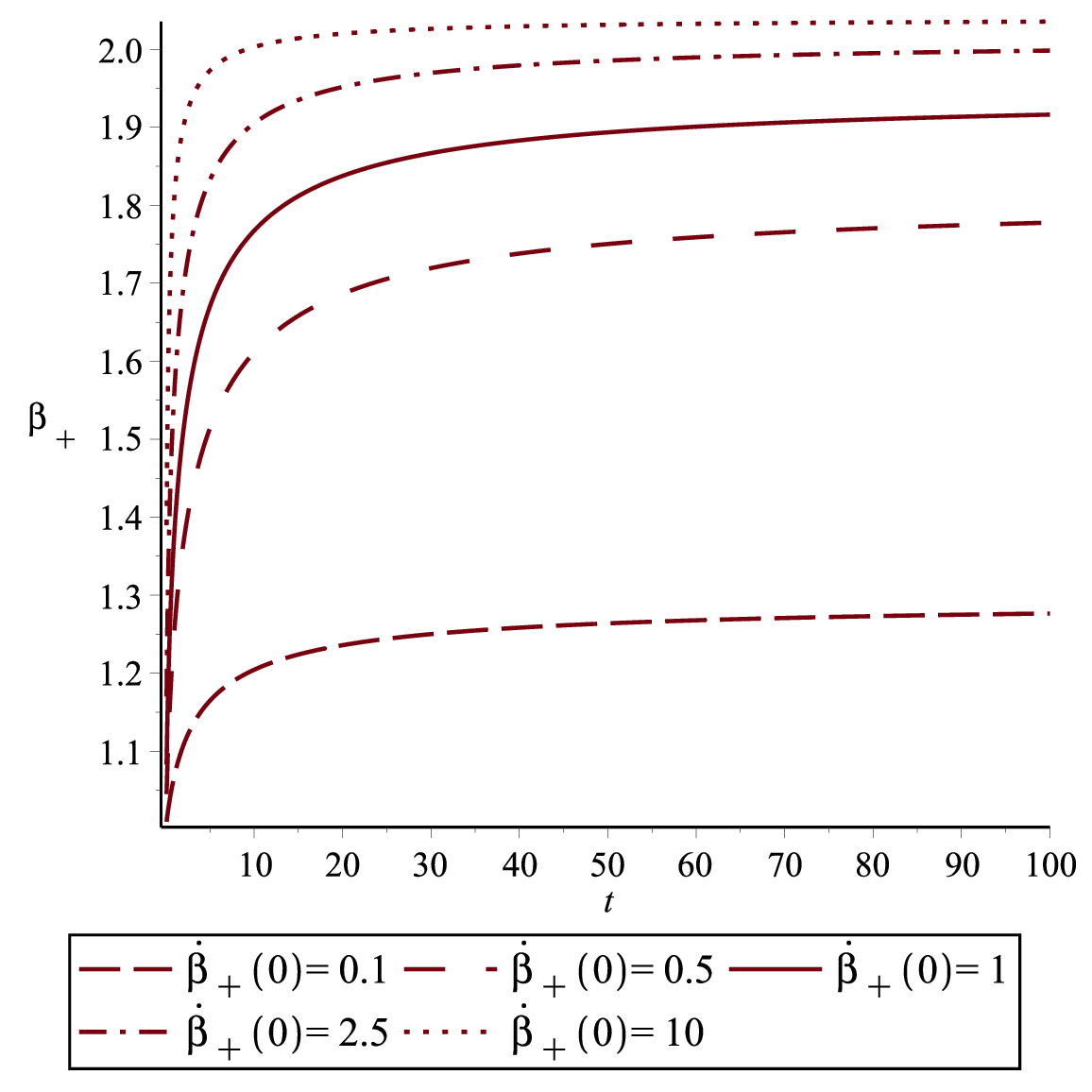}
		{\small Figure 12: Behavior of $\beta_{+} (t)$ for different values of $\dot{\beta}_{+} (0)$ = $\dot{\beta}_{-} (0)$. Here, we considered $C$ = 0.1, $C_-$ = $C_+$, and $\gamma$ = $- 0.5$.}
		\label{Bcomv1} 
	\end{minipage}\hfill
\end{figure} 

\begin{table}[!tbp]
	\centering
	\caption{\small Values of $a_{nc}$, $\dot{a}_{nc}$, $\beta_{+}$ and $\dot{\beta}_{+}$ for different values of $\dot{\beta}_{+} (0)$ = $\dot{\beta}_{-} (0)$ and $t$, with $C$ = 0.1, $C_-$ = $C_+$ and $\gamma$ = - 0.5.}
	\label{T6}
	{\tiny\begin{tabular}{c c c c c c} 
			\hline
			$\dot{\beta}_{+} (0)$ &  $t$ & $a_{nc} (t)$ & $\dot{a}_{nc} (t)$ & $\beta_{+} (t)$ & $\dot{\beta}_{+} (t)$ \\
			
			\hline

			$0.1$ & $10^{2}$ & $1.48593441474443 \times 10$ & $8.63074029050892 \times 10^{-2}$ & $1.27652561319558$ & $1.42152520665372 \times 10^{-4}$ \\
			
			$0.1$ & $10^{5}$ & $1.11784526346040 \times 10^{3}$ & $7.39052628309245 \times 10^{-3}$ & $1.29415710347157$ & $3.41459070805716 \times 10^{-10}$ \\
			
			$0.1$ & $10^{10}$ & $2.65938662039816 \times 10^{6}$ & $1.80749820251672 \times 10^{-4}$ & $1.29419427002694$ & $8.60589971258084 \times 10^{-20}$ \\
			
			$0.1$ & $10^{50}$ & $4.21090364232810 \times 10^{33}$ & $2.86341338252175 \times 10^{-17}$ & $1.29419427096240$ & $1.36620807030152 \times 10^{-96}$ \\ 
			
			$0.1$ & $10^{100}$ & $4.21084855453572 \times 10^{67}$ & $2.86337738814396 \times 10^{-33}$ & $1.29419427096240$ & $1.36909036648383 \times 10^{-192}$ \\ 
			
			\hline
			
			$0.5$ & $10^{2}$ & $1.72271075085962 \times 10$ & $1.04428656362462 \times 10^{-1}$ & $1.77775449161700$ & $2.99488806657740 \times 10^{-4}$ \\
			
			$0.5$ & $10^{5}$ & $1.49381487663321 \times 10^{3}$ & $1.00040186673963 \times 10^{-2}$ & $1.81283205979313$ & $6.15176342387333 \times 10^{-10}$ \\ 			
			
			$0.5$ & $10^{10}$ & $3.64395457941376 \times 10^{6}$ & $2.47724420704324 \times 10^{-4}$ & $1.81289895531879$ & $1.54706368715200 \times 10^{-19}$ \\ 
			
			$0.5$ & $10^{50}$ & $5.77237595965919 \times 10^{33}$ & $3.92521641774465 \times 10^{-17}$ & $1.81289895700041$ & $2.45602204897043 \times 10^{-96}$ \\    
			
			$0.5$ & $10^{100}$ & $5.77230119749356 \times 10^{67}$ & $3.92516495030003 \times 10^{-33}$ & $1.81289895700041$ & $2.46114686652164 \times 10^{-192}$ \\ 
			
			\hline

			$1$ & $10^{2}$ & $2.12552453354794 \times 10$ & $1.33574497782499 \times 10^{-1}$  & $1.91626231546279$ & $2.43518154705003 \times 10^{-4}$ \\
			
			$1$ & $10^{5}$ & $2.02266726475716 \times 10^{3}$ & $1.36416914112010 \times 10^{-2}$ & $1.94389881461940$ & $4.64068967741611 \times 10^{-10}$ \\ 
			
			$1$ & $10^{10}$ & $5.00170300783830 \times 10^{6}$ & $3.40068868934304 \times 10^{-4}$ & $1.94394926373599$ & $1.16629412774537 \times 10^{-19}$ \\  
			
			$1$ & $10^{50}$ & $7.92501170584300 \times 10^{33}$ & $5.38900869343990 \times 10^{-17}$ & $1.94394926500373$ & $1.85152591704734 \times 10^{-96}$ \\   
			
			$1$ & $10^{100}$ & $7.92490917376261 \times 10^{67}$ & $5.38893879841991 \times 10^{-33}$ & $1.94394926500373$ & $1.85540354115470 \times 10^{-192}$ \\ 
			
			\hline
			
			$2.5$ & $10^{2}$ & $3.27933895831971 \times 10$ & $2.14565404616915 \times 10^{-1}$ & $1.99836523596700$ & $1.35339151796672 \times 10^{-4}$ \\
			
			$2.5$ & $10^{5}$ & $3.38921358482079 \times 10^{3}$ & $2.29795199602926 \times 10^{-2}$ & $2.01334416625309$ & $2.44513875168261 \times 10^{-10}$ \\ 
			
			$2.5$ & $10^{10}$ & $8.46640412595575 \times 10^{6}$ & $5.75687505749010 \times 10^{-4}$ & $2.01337074447312$ & $6.14345810105080 \times 10^{-20}$ \\
			
			$2.5$ & $10^{50}$ & $1.34169835761710 \times 10^{34}$ & $9.12354863746835 \times 10^{-17}$ & $2.01337074514091$ & $9.75315671893138 \times 10^{-97}$ \\ 
			
			$2.5$ & $10^{100}$ & $1.34168081024688 \times 10^{68} $ & $9.12342574015682 \times 10^{-33}$ & $2.01337074514091$ & $9.77330574150987 \times 10^{-193}$ \\ 
			
			\hline
			
			$10$ & $10^{2}$ & $7.62169794077434 \times 10$ & $5.13076737540624 \times 10^{-1}$ & $2.03553949195842$ & $4.27364187272902 \times 10^{-5}$ \\
			
			$10$ & $10^{5}$ & $8.25011112420169 \times 10^{3}$ & $5.60725477006067 \times 10^{-2}$ & $2.04020132872402$ & $7.50318338503883 \times 10^{-11}$ \\  
			
			$10$ & $10^{10}$ & $2.07038699213482 \times 10^{7}$ & $1.40785207776121 \times 10^{-3}$ & $2.04020948429345$ & $1.88510584817603 \times 10^{-20}$ \\  
			
			$10$ & $10^{50}$ & $3.28125786722732 \times 10^{34}$ & $2.23125547542435 \times 10^{-16}$ & $2.04020948449835$ & $2.99264858893673 \times 10^{-97}$ \\ 
			
			$10$ & $10^{100}$ & $3.28121540077935 \times 10^{68} $ & $2.23122675543417 \times 10^{-32}$ & $2.04020948449835$ & $2.99893152725601 \times 10^{-193}$ \\  				
			
			\hline
	\end{tabular}}
\end{table}

	\subsection{\textbf{$\beta_{\pm} (0)$ variation}}

After our investigations on how the evolution of the NC scale factors are modified by the variation of $\beta_{\pm} (0)$, we notice that $a_{nc}$ has no significant dependence on $\beta_{\pm} (0)$. It means that, there is no significant differences between graphs of $a_{nc} \times t$, for different values of $\beta_{\pm} (0)$. An example of this result is shown in Table \ref{T7}. On the other hand, when studying how the variation of $\beta_{\pm} (0)$ influences the dynamics of $\beta_\pm$, we notice that for all values of $\beta_{\pm} (0)$, $\beta_\pm$ always goes to a constant value after a period of expansion. We cannot say which values of $\beta_{\pm} (0)$ make $\beta_\pm$ goes to a constant more rapidly or slowly, because there is not a significant difference between them.
In Figure 13, we present an example of that effect, in a graph of $\beta_{+} \times t$ (as explained, above, $\beta_{-}$ behaves in the same way), where we use five different values of $\beta_{+} (0)$ ($\beta_{+} (0)$ = 0.01, 0.025, 0.05, 0.075 and 0.1). In Table \ref{T7}, we present the values of the scale factors and their time derivatives for different times and $\beta_{+} (0)$ values. It is noted that the values of $\beta_{+} (t)$ stabilize while $\dot{\beta}_{+} (t)$ tend to zero, which characterizes the phenomenon of isotropization. On the other hand, $a_{nc} (t)$ is always expansive, although in a decelerated way, as shown by the values of $\dot{a}_{nc} (t)$.

\begin{figure}[!tbp]
	\begin{minipage}{0.4\textwidth}
		\centering
		\includegraphics[width=\linewidth]{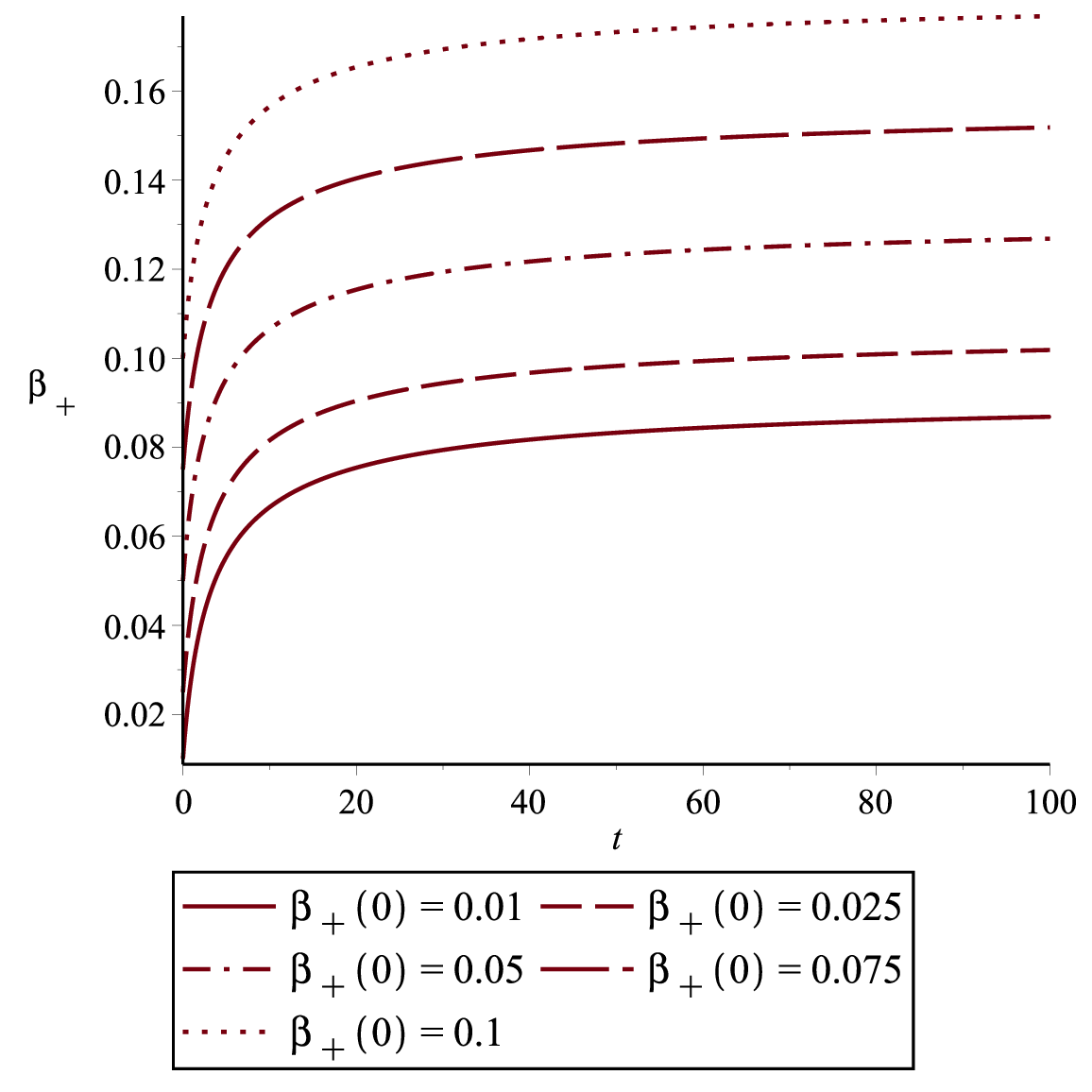}
		{\small Figure 13: Behavior of $\beta_{+} (t)$ for different values of $\beta_{+} (0)$. Here, we considered $C$ = $C_+$ = $C_-$ = 0.1, $\dot{\beta}_{\pm} (0) = 0.026352314$ and $\gamma$ = $- 0.5$.}
		\label{BcomB1}
	\end{minipage}\hfill
	\end{figure} 

\begin{table}[!tbp]
	\centering
	\caption{\small Values of $a_{nc}$, $\dot{a}_{nc}$, $\beta_+$ and $\dot{\beta}_{+}$ for different values of $\beta_+ (0)$ and $t$, with $C = C_+ = C_- = 0.1$, $\dot{\beta}_{\pm} (0) = 0.026352314$ and $\gamma = -0.5$.}
	\label{T7}
		{\tiny\begin{tabular}{c c c c c c}
				\hline
				$ \beta_+ (0) $ &  $t $ & $a_{nc} (t)$ & $\dot{a}_{nc} (t)$ & $\beta_{+} (t)$ & $\dot{\beta}_{+} (t)$ \\
				\hline

				0.01 & $10^{2}$ & $1.47261852880949 \times 10$ & $8.52059228078514\times 10^{-2}$ & $8.68444296484373\times 10^{-2}$ & $4.01812744911463\times 10^{-5}$ \\
				
				0.01 & $10^{5}$ & $1.09169179219927 \times 10^{3}$ & $7.20701312944106\times 10^{-3}$ & $9.18606914292346\times 10^{-2}$ & $9.83963360997730 \times 10^{-11}$ \\
				
				0.01 & $10^{10}$ & $2.58968686083657 \times 10^{6}$ & $1.76007894617209\times 10^{-4}$ & $9.18714030948241\times 10^{-2}$ & $2.48079440395924 \times 10^{-20}$ \\
				
				0.01 & $10^{50}$ & $4.10033281387534 \times 10^{33}$ & $2.78822596554165 \times 10^{-17}$ & $9.18714033644820\times 10^{-2}$ & $3.93830611927357 \times 10^{-97}$ \\
				
				0.01 & $10^{100}$ & $4.10027956069624 \times 10^{67}$ & $2.78819063589185 \times 10^{-33}$ & $9.18714033644821\times 10^{-2}$ & $3.94660953614205 \times 10^{-193}$ \\
				
				\hline
				
				0.025 & $10^{2}$ & $1.47261851380801 \times 10$ & $8.52059338457978 \times 10^{-2}$ & $1.01844434460444 \times 10^{-1}$ & $4.01813952378685 \times 10^{-5}$ \\
				
				0.025 & $10^{5}$ & $1.09169177547518 \times 10^{3}$ & $7.20701303195885\times 10^{-3}$ & $1.06860698331230 \times 10^{-1}$ & $ 9.83963733803563 \times 10^{-11}$ \\
				
				0.025 & $10^{10}$ & $2.58968678997633 \times 10^{6}$ & $1.76007852956498 \times 10^{-4}$ & $1.06871409999141 \times 10^{-1}$ & $2.48076669640442 \times 10^{-20}$ \\
				
				0.025 & $10^{50}$ & $4.10033289305991 \times 10^{33}$ & $2.78822536324384 \times 10^{-17}$ & $1.06871410268798 \times 10^{-1}$ & $3.93827439311357 \times 10^{-97}$ \\
				
				0.025 & $10^{100}$ & $4.10027929390874 \times 10^{67}$ & $2.78819034623523 \times 10^{-33}$ & $1.06871410268798 \times 10^{-1}$ & $3.94658704570934 \times 10^{-193}$ \\
				
				\hline
				
				0.05 & $10^{2}$ & $1.47261848139896 \times 10$ & $8.52059347226026 \times 10^{-2}$ & $1.26844438794072 \times 10^{-1}$ & $4.01814174638344 \times 10^{-5}$ \\
				
				0.05 & $10^{5}$ & $1.09169175929142 \times 10^{3}$ & $7.20701295360640 \times 10^{-3}$ & $1.31860704969430 \times 10^{-1}$ & $9.83964140568461 \times 10^{-11}$ \\
				
				0.05 & $10^{10}$ & $2.58968679812004 \times 10^{6}$ & $1.76007818343593 \times 10^{-4}$ & $1.31871416639692 \times 10^{-1}$ & $2.48074693361900 \times 10^{-20}$ \\
				
				0.05 & $10^{50}$ & $4.10033295869173 \times 10^{33}$ & $2.78822523947839 \times 10^{-17}$ & $1.31871416909349 \times 10^{-1}$ & $3.93827529069912 \times 10^{-97}$ \\
				
				0.05 & $10^{100}$ & $4.10027919239829 \times 10^{67}$ & $ 2.78818984532237 \times 10^{-33}$ & $1.31871416909349 \times 10^{-1}$ & $3.94655412368035  \times 10^{-193}$ \\
				
				\hline
				
				0.075 & $10^{2}$ & $1.47261847316966 \times 10$ & $8.52059345807472 \times 10^{-2}$ & $1.51844439679822 \times 10^{-1}$ & $4.01814179551821 \times 10^{-5}$ \\
				
				0.075 & $10^{5}$ & $1.09169175591580 \times 10^{3}$ & $7.20701293489404 \times 10^{-3}$ & $1.56860706385155 \times 10^{-1}$ & $9.83964209788674 \times 10^{-11}$ \\ 
				
				0.075 & $10^{10}$ & $2.58968680211751 \times 10^{6}$ & $1.76007813141873 \times 10^{-4}$ & $1.56871418055898 \times 10^{-1}$ & $2.48074434412398 \times 10^{-20}$ \\  
				
				0.075 & $10^{50}$ & $4.10033296508804 \times 10^{33}$ & $2.78822525402685 \times 10^{-17}$ & $1.56871418325556 \times 10^{-1}$ & $3.93827793662011 \times 10^{-97}$ \\
				
				0.075 & $10^{100}$ & $4.10027918315254 \times 10^{67}$ & $2.78818974557760 \times 10^{-33}$ & $1.56871418325556 \times 10^{-1}$ & $3.94654805491996 \times 10^{-193}$ \\
				
				\hline
				
				0.1 & $10^{2}$ & $1.47261847316966 \times 10$ & $8.52059345807472 \times 10^{-2}$ & $1.76844439679822 \times 10^{-1}$ & $4.01814179551821 \times 10^{-5}$ \\
				
				0.1 & $10^{5}$ & $1.09169175591580 \times 10^{3}$ & $7.20701293489404 \times 10^{-3}$ & $1.81860706385155 \times 10^{-1}$ & $9.83964209788674 \times 10^{-11}$ \\			
				
				0.1 & $10^{10}$ & $2.58968680211751 \times 10^{6}$ & $1.76007813141873\times 10^{-4}$ & $1.81871418055898 \times 10^{-1}$ & $2.48074434412398 \times 10^{-20}$ \\  
				
				0.1 & $10^{50}$ & $4.10033296508804 \times 10^{33}$ & $2.78822525402685 \times 10^{-17}$ & $1.81871418325556 \times 10^{-1}$ & $3.93827793662011 \times 10^{-97}$ \\ 
				
				0.1 & $10^{100}$ & $4.10027918315254 \times 10^{67}$ & $2.78818974557760 \times 10^{-33}$ & $1.81871418325556 \times 10^{-1}$ & $3.94654805491996 \times 10^{-193}$ \\
				
				\hline
		\end{tabular}}
\end{table}

\section{Estimates for the value of the NC parameter}
\label{estimativa}

In the present section, we want to estimate the value of $\gamma$, using the present observational data. In order to do this, we start with the NC Hamiltonian Eq. (\ref{NH-NC}), in the gauge $N_{nc} = 1$. We, then, impose some conditions, in the general NC Hamiltonian, such that it may better describe the present properties of our universe. Therefore, we consider that the isotropization has already occurred ($p_{{\pm}_{nc}} \rightarrow 0 $) and the accelerated expansion is due, entirely, to the presence of the NC parameter. Under those conditions, the NC Hamiltonian Eq. (\ref{NH-NC}) reduces to,
\begin{equation}
	\mathcal{H}_{nc}  = - \frac{{p_{{a}_{nc}}^2}}{24 a_{nc}} + {a_{nc}}^{- 3 \omega}  p_{{T}_{nc}} \, .
\label{H red}
\end{equation}
Unlike the previous sections, here, we work with the NC variables and, therefore, we do not introduce the transformations Eq. (\ref{parentes1})-(\ref{parentes3}). This will allow us to find the NC isotropic scale factor $a_{nc}$ as a function of time.

Next, from the above NC Hamiltonian Eq. (\ref{H red}), we compute the dynamical equations with the aid of the deformed Poisson brackets 
Eqs. (\ref{PBs}). They are given by,
\begin{equation}
	\dot{T}_{nc} = \left \{T_{nc}, H_{nc} \right \} = 0 \, , 
	\label{PBnc3}
\end{equation}
\begin{equation}
	\dot{a}_{nc} = \left \{ a_{nc} , H_{nc} \right \} = - \frac{1}{12} \frac{p_{{a}_{nc}}}{a_{nc}} + \gamma \,  {a_{nc}}^{- 3 \omega} \, ,
	\label{PBnc1}
\end{equation}
\begin{equation}
		\dot{p}_{{a}_{nc}} = \left \{{p}_{{a}_{nc}}, H_{nc} \right \} = - \frac{p^{2}_{a_{nc}}}{24 a^{2}_{nc}} + \frac{3 \omega \, p_{T_{nc}}}{a^{3 \omega + 1}_{nc}}\, ,
		\label{PBnc2} 
\end{equation}
\begin{equation}
	\dot{p}_{{T}_{nc}} = \left \{{p}_{{T}_{nc}}, H_{nc} \right \} = \gamma \, \left [ - \frac{p^{2}_{a_{nc}}}{24 a^{2}_{nc}} + \frac{3 \omega \, p_{T_{nc}}}{a^{3 \omega + 1}_{nc}}\right ] \, .
	\label{PBnc4}
\end{equation} 

We want to write a dynamical equation for the NC isotropic scale factor. So, starting from Eq. (\ref{PBnc1}) we can find an expression to $p_{{a}_{nc}}$, 
\begin{equation}
	p_{{a}_{nc}} = - 12 \, a_{nc} \, (\dot{a}_{nc} - \gamma \, a^{-3 \omega}_{nc}) \, ,  
	\label{pAnc}
\end{equation} 
We can also see from Eqs. (\ref{PBnc2}) and (\ref{PBnc4}), that,
\begin{equation}
	\dot{p}_{T_{nc}} = \gamma \, \dot{p}_{a_{nc}} \, ,
\end{equation}
that can be integrated and provides,
\begin{equation}
	{p}_{T_{nc}} = \gamma \, {p}_{a_{nc}} + C_f \, ,
	\label{pTnc}
\end{equation}
where $C_f$ is a constant associated to the fluid energy density. 

In order to estimate the NC parameter $\gamma$ we set $\mathcal{H}_{nc}$ Eq. (\ref{H red}) equal to zero, so that,
\begin{equation}
	- \frac{{p_{{a}_{nc}}^2}}{24 a_{nc}} + {a_{nc}}^{- 3 \omega}  p_{{T}_{nc}}  = 0 \, .
	\end{equation}
Substituting Eqs. (\ref{pAnc}) and (\ref{pTnc}) in the above equation, keeping up to the second order in $\gamma$, we obtain the following equation for the time evolution of $a_{nc}$,
\begin{equation}
	\dot{a}^{2}_{nc} - {\gamma}^{2} {a^{-6 \omega}_{nc}} - \frac{C_f}{6} a^{- 3 \omega - 1}_{nc} = 0, 
	\label{a-nc-gama}
\end{equation}
Now, we assume that the matter content of the model consists of dust ($\omega$ = 0) and that the accelerated expansion is due entirely to the presence of the NC parameter $\gamma$. Doing this, we obtain that Eq. (\ref{a-nc-gama}), simplifies to,
\begin{equation}
	\dot{a}_{nc} (t) = \left ({\gamma}^{2}  + \frac{C_f}{6 \cdot  a_{nc} (t)} \right )^{\frac{1}{2}} \, .
	\label{a-nc-dot}
\end{equation}
This equation (\ref{a-nc-dot}) has a general solution given by,
\begin{equation}
	t + \frac{1}{72} \, \, \frac{\sqrt{6} \, \xi \left [ C_f \, \ln(\frac{A + B \gamma}{12 \, \gamma})\sqrt{6} - B \right ]}{\xi \, \gamma^3} + D = 0 , 
	\label{eq-estimativa}
\end{equation}
where $\xi = 6  \, \gamma^2 \, a_{nc}(t) + C_f$, $A =  \sqrt{6} \, (12 \,  \gamma ^2 \, a_{nc} (t) + C_f)$, $B = 12 \, \sqrt{a_{nc} (t) \, \xi}$ and $D$ is an integration constant. The constant $D$, can be determined by providing an initial condition $a_{nc} (t = t_0) \equiv a_0$. Since the matter content of our model is dust, we shall consider $t_0$ and $a_0$ as the time and the scale factor at the epoch of matter-radiation equality. After that epoch, matter dominates the Universe. In approximated values, one has \cite{liddle2003},
\begin{equation}
	t_0 \approx 3.5 \times 10^5, \,\,\, a_0 \approx 2.84 \times 10^{-4}.
	\label{condiçãoinicial}
\end{equation} 
After determining the value of $D$ Eq. (\ref{eq-estimativa}) with the aid of the initial conditions Eqs. (\ref{condiçãoinicial}), we notice that
$D$ becomes a very complicated function of $C_f$ and $\gamma$, $D = D(C_f, \gamma)$. Introducing that result in Eq. (\ref{eq-estimativa}), we obtain,
\begin{equation}
	t + \frac{1}{72} \, \, \frac{\sqrt{6} \, \xi \left [ C_f \, \ln(\frac{A + B \gamma}{12 \, \gamma})\sqrt{6} - B \right ]}{\xi \, \gamma^3} + D(C_f, \gamma) = 0 , 
	\label{eq-estimativaD}
\end{equation}
where $D(C_f, \gamma)$ is a very complicated function of $C_f$ and $\gamma$. 

In order to estimate the value of the NC parameter $\gamma$, from Eq. (\ref{eq-estimativaD}), we must furnish the values of the variables $a_{nc}(t)$ and $t$ and the constant $C_f$. The constant $C_f$, which is
associated to the dust energy density, can be written in terms of two important observable quantities: the present mass density parameter ($\Omega_{m0}$) and the present Hubble constant ($H_{0}$). It is given by: $C_f = \Omega_{m0} H_{0}^2$ \cite{liddle2003}. Here, we use the following values for these two quantities: $\Omega_{m0} = 0.3089$ and $H_{0} = 67.74 (km/s)/Mpc$ \cite{ade2016}. Next, we must provide
the values of the pair $a_{nc}(t)$ and $t$. We choose these values to represent the time ($t^{ae}$) and scale factor ($a_{nc}^{ae}$) when the present accelerated expansion of the Universe began.
More precisely, we consider ten values of these quantities that produce ten estimated values of $\gamma$ as solutions to Eq. (\ref{eq-estimativaD}). The values of $a_{nc}^{ae}$ and $t^{ae}$ are shown in the first two columns of Table \ref{T8}. For a given value of $a_{nc}^{ae}$, the corresponding value of $t^{ae}$ was obtained using the \emph{Cosmological Calculator for the Flat Universe} \cite{gnedin}, considering the values of $\Omega_{m0} $ and $H_{0}$ given above. Therefore, the numerical solution to Eq. (\ref{eq-estimativaD}), considering the assumed value of $C_f$, allows us to obtain the ten values of $\gamma$ given in the third column of Table \ref{T8}. The last column of this Table is the numerical error involved in computing $\gamma$, numerically, from Eq. (\ref{eq-estimativaD}). From Table \ref{T8}, one notices that $\gamma$ increases when $a_{nc}^{ae}$ and $t^{ae}$ decrease toward the initial moments of the Universe. This result is compatible with what would be expected, since it is expected that NCTY must have been more important at the beginning of the universe.

\begin{table}[!tbp]
	\caption{\small Table with ten different estimated values of $\gamma$.}
	\label{T8}
	{\small\begin{tabular}{c c c c}
			\hline
			$a^{ae}_{nc}$ &  $t^{ae}$ ($10^{9}$ years) & $\gamma$ ($10^{-18}$) & Numerical error (\%)   \\
			\hline
			
			$1.0$ & $13.7915$ & $2.082712384$ & $9.195742436 \times 10^{-8}$\\
			
			$0.9$ & $12.3111$ & $2.083105569$ & $0$ \\ 
			
			$0.8$ & $10.7588$ & $2.101554930$ & $8.840884797 \times 10^{-8}$ \\ 
			
			$0.7$ & $9.1480$ & $2.145356273$ & $0$\\    
			
			$0.6$ & $7.5030$ & $2.225406976$ & $4.225745762 \times 10^{-8}$ \\    
			
			$0.5$ & $5.8623$ & $2.358404433$ & $1.081683655 \times 10^{-7}$  \\ 
			
			$0.4$ & $4.2784$ & $2.572455458$ & $0$ \\ 
			
			$0.3$ & $2.8148$ & $2.922858692$ & $1.126395142 \times 10^{-7}$ \\    
			
			$0.2$ & $1.5428$ & $3.548830601$ & $2.055079754 \times 10^{-7}$ \\    
			
			$0.1$ & $0.5469$ & $5.003208668$ & $ 2.318944630 \times 10^{-7}$ \\

			\hline
	\end{tabular}}
\end{table}

\section{The non-commutative Bianchi-I model for different fluids}
\label{fluidos}

In Section \ref{NCradiation}, we made a detailed study of the NC BI model for the case of a radiation perfect fluid. However, it is possible to study the model coupled to different types of perfect fluids, that is, to different values of the constant $\omega$ Eq. (\ref{eqestado}). In the present section, we want to show how the presence of different types of perfect fluids modify the dynamics of the scale factors $a_{nc}$ and $\beta_\pm$. It means that, we shall not make, here, a detailed study of each NC BI model coupled to different types of perfect fluids as we did, for the case of radiation, in Section \ref{NCradiation}. We restrict our attention to the following 
perfect fluids: cosmological constant ($\omega$ = -1), domain walls ($\omega$ = -2 /3), cosmic strings ($\omega$ = -1/3), radiation ($\omega$ = 1/3), dust ($\omega$ = 0) and stiff matter ($\omega$ = 1).

For each type of perfect fluid, we start writing the appropriated system of differential equations derived from Eqs. (\ref{T ponto1}), (\ref{evoA}), (\ref{evoBeta}), (\ref{tipo friedmann}) and (\ref{evoluçãobetas}), by the use of the correct $\omega$. Then, we solve each system of differential equations for many different values of the free parameters and initial conditions and find $a_{nc}(t)$ and $\beta_\pm(t)$, for each fluid. After comparing the dynamical evolution of $a_{nc}(t)$ between the different fluids, we notice that $a_{nc}(t)$ expands more rapidly
for fluids with smaller values of $\omega$. An example of that comparison is shown in Figure 14. Now, comparing the dynamical evolution of $\beta_\pm(t)$ between the different fluids, we notice that the isotropization of the model occurs more rapidly for  fluids with smaller values of $\omega$. An example of that comparison is shown in Figure 15. Here, we imposed the same conditions as in the radiation case concerning the anisotropic scale factors $\beta_\pm$. It means that, the graph for the evolution of $\beta_{-}$ has a similar behavior to that of $\beta_{+}$. Therefore, it was omitted. In both Figures 14 and 15, we considered: $\gamma = 10^{-18}$, $C$ = $C_+$ = $C_-$ = 0.1, $T_{c} (0)$ = 0 and $a_{c} (0)$ = $\beta_{+} (0) $ = $\beta_{-} (0)$ = 1.

\begin{figure}[!tbp]
	\begin{minipage}{0.4\textwidth}
		\centering
		\includegraphics[width=\linewidth]{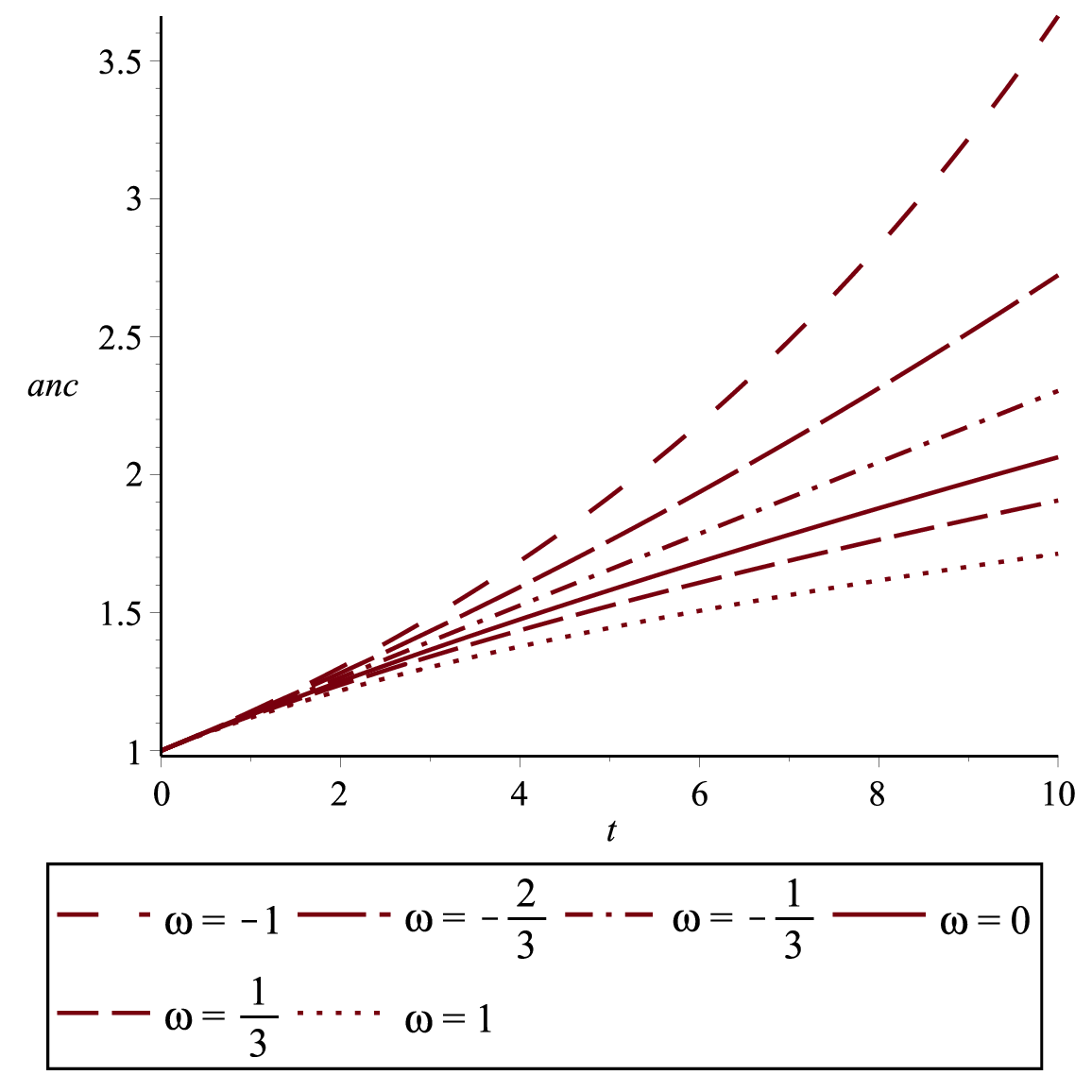}
		{\small Figure 14: Behavior of $a_{nc} (t)$ for different values of $\omega$ for the non-commutative case.}
		\label{fig14}
	\end{minipage}\hfill
	\begin{minipage}{0.4\textwidth}
		\centering
		\includegraphics[width=\linewidth]{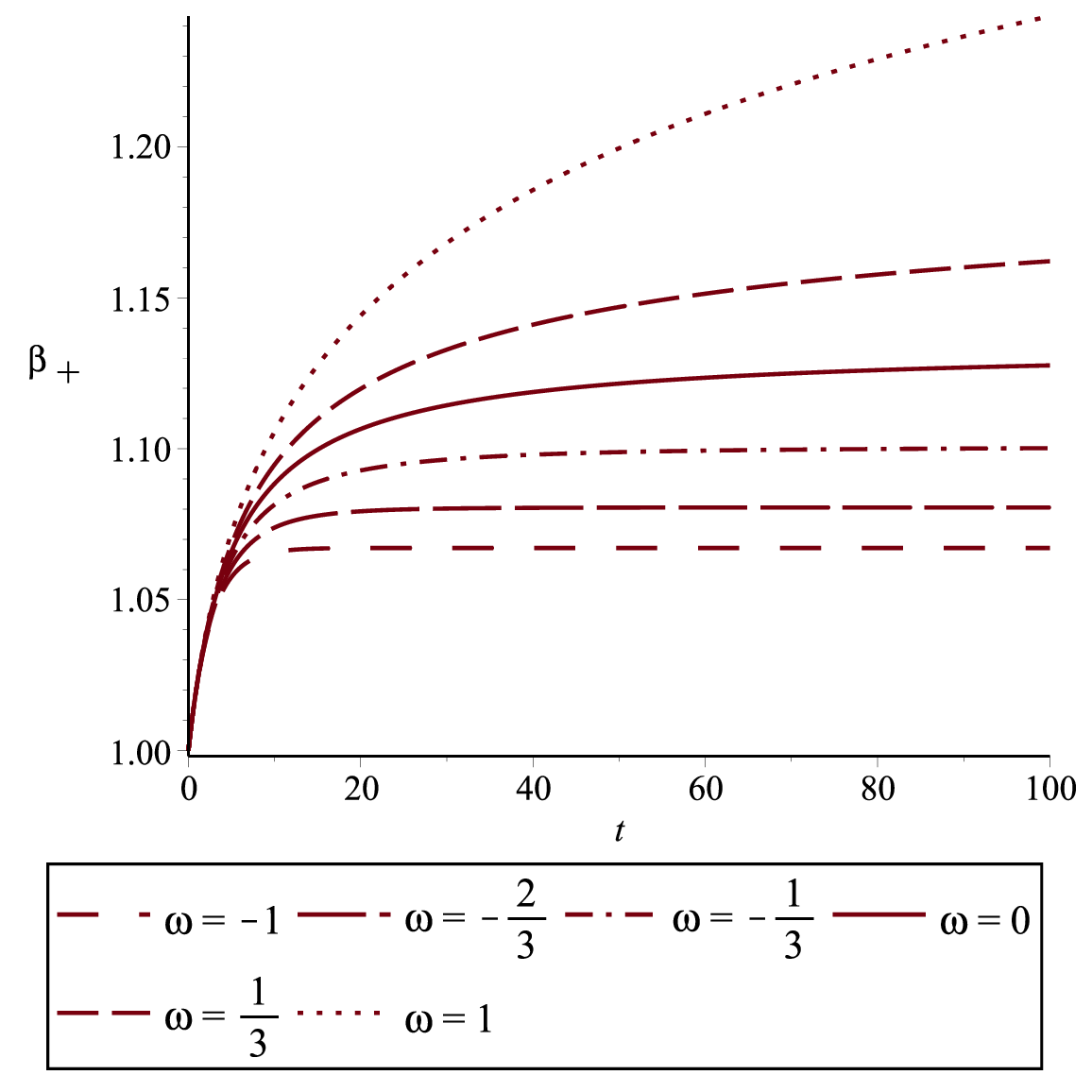}
		{\small Figure 15: Behavior of $\beta_{+} (t)$ for different values of $\omega$ for the non-commutative case.}
		\label{fig15}
	\end{minipage}\hfill
\end{figure}

\section{Conclusions}	 
\label{conclusoes}

In the present work, we studied the dynamical evolution of an homogeneous and anisotropic, NC BI model coupled to a radiation perfect fluid. We considered general relativity as the gravitational theory. Our first motivation was determining if the present model tends to an homogeneous and isotropic NC FRW model, during its evolution. More precisely, we wanted to establish how the parameters and initial conditions of the model, quantitatively, influence the isotropization of the present model. In order to simplify our task, we used the Misner parametrization of the BI metric. In terms of that parametrization the BI metric has three metric functions: the isotropic scale factor $a(t)$ and the anisotropic scale factors $\beta_\pm (t)$, which measure the spatial anisotropy of the model. Our second motivation was trying to explain the present accelerated expansion of the universe by means of NCTY. We started by writing the total Hamiltonian of the model, which is a sum of the gravitational Hamiltonian with the matter Hamiltonian. Initially, we considered the matter Hamiltonian for a generic perfect fluid. The NCTY was introduced by two nontrivial Poisson brackets between some geometrical as well as matter variables of the model. We associated the same NC parameter $\gamma$ to both nontrivial Poisson brackets. We recovered the description in terms of commutative variables by introducing some variables transformations that depend on the NC parameter. Using those variables transformations, we rewrote the total NC Hamiltonian of the model in terms of commutative variables. From the resulting Hamiltonian, we obtained the dynamical equations for a generic perfect fluid. In order to solve these equations, we restricted our attention to a model where the perfect fluid is radiation. The solutions for $a_{nc} (t)$ are forever expansive and the solutions for $\beta_\pm (t)$ start expanding and soon tend to constant values. They all depend on four parameters: $\gamma$, a parameter associated with the fluid energy ($C$) and two parameters which have no clear physical interpretation ($C_\pm$). They also depend on the initial conditions of the model variables: $a_c (0)$, $\dot{a}_{c} (0)$, $\beta_{\pm} (0)$ and $\dot{\beta}_{\pm} (0)$. 
With respect to the variation of the four parameters, we obtained the following results. (i) For larger values of the modulus of $\gamma$ the expansion rate increases and the time required for isotropization diminishes. These effects are more intense for negative values of $\gamma$. 
(ii) The energy density parameter $C$ is positive and the higher its value, the faster the expansion becomes and the shorter the time for isotropization. (iii) The anisotropy parameters $C_+$ and $C_-$ alter the solutions in the following way: larger $C_\pm$ values result in higher expansion rates. It does not seem to have a clear pattern on how the values of $C_\pm$ influence how fast $\beta_\pm$ goes to a constant value. (iv) The model tends to a FRW model compatible with the current universe.
With respect to the variation of the four initial conditions, we obtained the following results. (i) If one increases the values of $\dot{a}_{c} (0)$ or $\dot{\beta}_{\pm} (0)$ or decreases the values of $a_c (0)$, $a_{nc} (t)$ expands more rapidly. (ii) The dynamics of $a_{nc} (t)$ is not significantly altered by the variation of $\beta_{\pm} (0)$. (iii) If one increases the values of $\dot{a}_{c} (0)$ or $\dot{\beta}_{\pm} (0) \geq 0$ or decreases the values of $a_c (0)$ or $\dot{\beta}_{\pm} (0) < 0$, the time required for isotropization diminishes. (iv) There is no significant differences on how $\beta_\pm$ goes to constant values, for different values of $\beta_{\pm} (0)$.
We compared the NC solutions to the corresponding commutative ones. 
The comparison showed that the introduction of NCTY increases the expansion rate of the isotropic scale factor, in this sense it may be considered as a possible candidate for describing the accelerated expansion of the universe. We obtained estimates for the NC parameter. The estimated values of $\gamma$ behave consistently, that is, for times closer to the beginning of the universe the higher the values of this NC parameter. This agrees with our assumption that NCTY should be more relevant at the beginning of the universe. Finally, we compared some results of the NC BI model coupled to radiation with the same NC BI model coupled to other perfect fluids. In particular, we
obtained that $a_{nc}(t)$ expands more rapidly and the isotropization of the model occurs more rapidly for fluids with smaller values of $\omega$. As our main result, we showed that the solutions, after some time, produce an isotropic universe and the presence of NCTY reduces the time required for isotropization. Based on that result, we can speculate that the solutions may represent an initial anisotropic stage of our Universe, that later, due to the expansion, became isotropic.

{\bf Acknowledgments}. T. M. Abreu thanks Programa de Pós-Graduação em Física (PPG-Física), Universidade Federal de Juiz de Fora (UFJF) for some financial support.

{\bf Data availability statement}. All data generated or analyzed during this study are included in this published article.



\begin{thebibliography}{refs}

\bibitem{wheeler} J. A. Wheeler, in {\it Batelles Rencontres}, eds. C. DeWitt
and J. A. Wheeler (Benjamin, New York, 1968), 242.

\bibitem{bianchi} L. Bianchi, `Sugli spazii a tre dimensioni che ammettono un gruppo continuo di movimenti',
Ital. Sci. Mem. di Mat. {\bf 11}, 267 (1898); a reprint of this paper, in English, was published in
Gen. Rel. Grav. {\bf 33}, 2171 (2001).

\bibitem{kasner} E. Kasner, `Geometrical theorems on Einstein's cosmological equations', Am. J. Math.
{\bf 43}, 217 (1921).

\bibitem{george} G. Efstathiou and S. Gratton, `The evidence for a spatially flat Universe', MNRAS {\bf 496}, L91-L95 (2020).

\bibitem{anton} A. Chudaykin, K. Dolgikh and M. M. Ivanov, `Constraints on the curvature of the Universe and dynamical dark energy from the full-shape and BAO data', Phys. Rev. D {\bf 103}, 023507 (2021).

\bibitem{jacobs} K. C. Jacobs, `Cosmologies of Bianchi type I with an uniform magnetic field', Ap. J. {\bf 155}, 379 (1969).

\bibitem{jacobs1} K. C. Jacobs, `Homogeneous electromagnetic and massive vector meson field in Bianchi cosmologies', Ap. J. {\bf 160}, 147 (1970).

\bibitem{banerjee} A. Banerjee, S. B. Duttachoudhury and A. K. Sanyal, `Bianchi type I cosmological model with a viscous fluid', J. Math. Phys. {\bf 26}, 3010–3015 (1985).

\bibitem{biesiada} M. Biesiada and M. Szydlowski, `Null geodesics in multi-dimensional homogeneous cosmologies', Phys. Lett. B {\bf 211}, 42 (1988).

\bibitem{chakraborty} S. Chakraborty, `Bianchi I cosmological model and the no-boundary condition', Int. J. Theor. Phys. {\bf 31}, 303 (1992). 

\bibitem{eath} P. D. D'Eath, `Quantization of the supersymmetric Bianchi-I model with a cosmological constant', Phys. Lett. B {\bf 320}, 12-15 (1994).

\bibitem{alvarenga} F. G. Alvarenga, A. B. Batista, J. C. Fabris and S. V. B. Gon\c{c}alves, `Troubles with quantum anisotropic cosmological
models: loss of unitarity', Gen. Relativ. Gravit. {\bf 35}, 1659 (2003).

\bibitem{pradhan} A. Pradhan and O. P. Pandey, `Bianchi type I anisotropic magnetized cosmological models with varying $\Lambda$', Int. J. Mod. Phys. D {\bf 12}, 1299-1314 (2003).

\bibitem{balakin} A. B. Balakin and W. Zimdahl, `Anisotropic cosmological models with non-minimally coupled magnetic field', Phys. Rev. D {\bf 71}, 124014 (2005).

\bibitem{balakin1} A. B. Balakin, `Magnetic relaxation in the Bianchi-I universe', Class. Quantum Grav. {\bf 24}, 5221–5245 (2007).

\bibitem{yadav} A. K. Yadav, F. Rahaman, S. Ray, and G. K. Goswami, `Magnetized dark energy and the late time acceleration', Eur. Phys. J. Plus {\bf 127}, 127 (2012).

\bibitem{canfora} F. Canfora, A. Giacomini and S. A. Pavluchenko, `Cosmological dynamics of gravitating hadron matter', Phys. Rev. D {\bf 90} 043516 (2014).

\bibitem{kamenshchik} A. Y. Kamenshchik, E. O. Pozdeeva, S. Y. Vernov, A. Tronconi and G. Venturi, `Bianchi-I cosmological model and crossing singularities', Phys. Rev. D {\bf 95}, 083503 (2017).

\bibitem{kamenshchik1} A. Y. Kamenshchik, E. O. Pozdeeva, S. Y. Vernov, A. A. Starobinsky, A. Tronconi and G. Venturi, `Induced gravity and minimally and conformally coupled scalar fields in Bianchi-I cosmological models', Phys. Rev. D {\bf 97}, 023536 (2018).

\bibitem{mandal} R. Mandal, S. Gangopadhyay and A. Lahiri, `Cosmology of Bianchi type-I metric using renormalization group approach for quantum gravity', Class. Quantum Grav. {\bf 37}, 065012 (2020).

\bibitem{parnovsky} S. L. Parnovsky, `The Big Bang could be anisotropic. The case of Bianchi I model', Class. Quantum Grav. {\bf 40}, 135005 (2023).

\bibitem{casadio} R. Casadio, A. Y. Kamenshchik, P. Petriakova and P. Mavrogiannis, `Bianchi cosmologies, magnetic fields and singularities', Phys. Rev. D {\bf 108}, 084059 (2023).

\bibitem{MTW} C. W. Misner, K. S. Thorne and J. A. Wheeler, {\it Gravitation}, (W. H. Freeman and Company, New York, 1973).

\bibitem{riess1998} Riess, Adam G., et al., `Observational evidence from supernovae for an accelerating universe and a cosmological constant', The astronomical journal {\bf 116} 1009 (1998).

\bibitem{perlmutter1999} Perlmutter, Saul, et al., `Measurements of $\Omega$ and $\Lambda$ from 42 high-redshift supernovae', The Astrophysical Journal {\bf 517} 565, (1999).

\bibitem{Mli} M. Li, X. D. Li, S. Wang and S. Wang, `Dark Energy', Commun. Theor. Phys. {\bf 56}, 525604 (2011).

\bibitem{trodden} A. Silvestri and M. Trodden, `Approaches to understanding cosmic acceleration', Rep. Prog. Phys. {\bf 72}, 096901 (2009).

\bibitem{vakili} B. Vakili, P. Pedram and S. Jalalzadeh, `Late time acceleration in a deformed phase space model of dilaton cosmology',
Phys. Lett. B {\bf 687}, 119–123 (2010).

\bibitem{el} A. R. El-Nabulsi, `Noncommutative accelerated multidimensional universe
dominated by quintessence', Astrophys. Space Sci. {\bf 326}, 163–167 (2010).

\bibitem{obregon} O. Obregon and I. Quiros, `Can noncommutative effects account for the present speed up of the cosmic expansion?',
Phys. Rev. D {\bf 84}, 044005 (2011).

\bibitem{gil} E. M. C. Abreu, M. V. Marcial, A. C. R. Mendes, W. Oliveira and G. Oliveira-Neto,
`Noncommutative cosmological models coupled to a perfect fluid and a cosmological constant',
JHEP {\bf 05}, p. 144 (2012).

\bibitem{vakili1} B. Malekolkalami, K. Atazadeh, B. Vakili, `Late time acceleration in a non-commutative model of modified cosmology',
Phys. Lett. B {\bf 739}, 400–404 (2014).

\bibitem{sabido} M. Sabido and C. Yee-Romero, `Deformed phase space Kaluza-Klein cosmology and late time acceleration', 
Phys. Lett. B {\bf 757}, p. 57 (2016).

\bibitem{gil1} G. A. Monerat, E. V. Corr\^{e}a Silva, C. Neves, G. Oliveira-Neto, 
L. G. Rezende Rodrigues and M. Silva de Oliveira, `Can noncommutativity affect the whole history
of the universe?', Int. J. Mod. Phys. D {\bf 26}, 1750022 (2017).

\bibitem{gil2} G. Oliveira-Neto and A. R. Vaz, `Noncommutative cosmological model in the presence
of a phantom fluid', Eur. Phys. J. Plus {\bf 132}, p. 131 (2017).

\bibitem{pourhassan} J. Sadeghi, B. Pourhassan, Z. Nekouee and M. Shokri, `Deformation of the quintom cosmological model and its consequences', 
Int. J. Mod. Phys. D {\bf 27}, 1850025 (2018).

\bibitem{gil3} E. M. C. Abreu, A. C. R. Mendes, G. Oliveira-Neto, J. Ananias Neto, L. G. Rezende
Rodrigues and M. Silva de Oliveira, `Horava–Lifshitz cosmological models with
noncommutative phase space variables', Gen. Relativ. Gravit. {\bf 51}, p. 95 (2019).

\bibitem{gil4} G. Oliveira-Neto and L. G. Rezende Rodrigues, `Noncommutative cosmological models induced
by a symplectic formalism coupled to phantom fluids', Int. J. Mod. Phys. A {\bf 34}, 1950206 (2019).

\bibitem{gil5} G. Oliveira-Neto and L. Fazza Marcon, `Complete noncommutativity in a cosmological model with radiation', 
Eur. Phys. J. Plus {\bf 136}, 584 (2021).

\bibitem{garcia} H. Garcia-Compean, O. Obregon and C. Ramirez, `Noncommutative Quantum Cosmology', Phys. Rev. Lett. {\bf 88}, 161301 (2002).

\bibitem{nelson} G. D. Barbosa and N. Pinto-Neto, `Noncommutative geometry and cosmology', Phys. Rev. D {\bf 70}, 103512(2004).

\bibitem{barbosa} G. D. Barbosa, `Noncommutative conformally coupled scalar field cosmology and its commutative counterpart', Phys. Rev. D {\bf 71}, 063511 (2005).

\bibitem{sabido1} W. Guzm\'{a}n, M. Sabido and J. Socorro, `Noncommutativity and scalar field cosmology', Phys. Rev. D {\bf 76}, 087302 (2007).

\bibitem{sabido2} W. Guzm\'{a}n, C. Ortiz, M. Sabido, J. Socorro and M. A. Aguero, `Noncommutative Bianchi Quantum Cosmology', Int. J. Mod. Phys. D {\bf 16}, 1625 (2007).

\bibitem{sabido3} C. Ortiz, E. Mena, M. Sabido and J. Socorro, `(Non)commutative Isotropization in Bianchi I with Barotropic Perfect Fluid and $\Lambda$ Cosmological', Int. J. Theor. Phys. {\bf 47}, 1240-1251 (2008).

\bibitem{socorro} J. Socorro, L. O. Pimentel, C. Ortiz and M. Aguero, `Scalar Field in the Bianchi I: Noncommutative Classical and Quantum Cosmology', Int. J. Theor. Phys. 48, 3567-3585 (2009).

\bibitem{gil6} G. Oliveira-Neto, M. Silva de Oliveira, G. A. Monerat and E. V. Corr\^{e}a Silva, `Noncommutativity in the early universe', Int. J. Mod. Phys. D {\bf 26}, 1750011 (2017).

\bibitem{moniz} S. M. M. Rasouli, N. Saba, M. Farhoudi, J. Marto and P.V. Moniz, `Inflationary universe in deformed phase space scenario', Ann. of Phys. {\bf 393}, 288 (2018).

\bibitem{saba} N. Saba and M. Farhoudi, `Noncommutative universe and chameleon field dynamics', Ann. of Phys. {\bf 395}, 1 (2018).

\bibitem{ghosh} P. Das and S. Ghosh, `Backreaction inhomogeneities in the cosmological parameter evolution via a noncommutative fluid', Phys. Rev. D {\bf 98}, 084047 (2018).

\bibitem{ghosh1} A. Krishna Mitra, R. Banerjee and S. Ghosh, `Noncommutative fluid and growing modes of inhomogeneity in (Newtonian) cosmology', JCAP {\bf 10}, 57 (2018).

\bibitem{bodmann} B. Bodmann, D. Hadjimichef, P. O. Hess, J. F. Pacheco, F. Weber, M. Razeira, G. A. Degrazia, M. Marzola and C. A. Z. Vasconcellos, `A Wheeler-DeWitt Non-Commutative Quantum Approach to the Branch-Cut Gravity', Universe  {\bf 9}, 428 (2023).

\bibitem{barron} L. R. D. Barron, A. E. Garcia, S. P. Payan and J. Socorro, `Is a loopy and noncommutative early Universe viable?', Phys. Lett. B {\bf 847}, 138299 (2023).

\bibitem{misner} C.W. Misner, `The Isotropy of the universe', Astrophys. J. {\bf 151}, 431 (1968).

\bibitem{schutz1970} B. F. Schutz, `Perfect fluids in general relativity: velocity potentials and a variational principle', Physical Review D {\bf 2}, 2762 (1970).

\bibitem{lapchinskii1977} Lapchinskii, V. G., Rubakov V. A., `Quantum gravitation: Quantization of the Friedmann model', Teoreticheskaya i Matematicheskaya Fizika {\bf 33}, 364-376 (1977).

\bibitem{liddle2003} A. Liddle, {\it An introduction to modern cosmology}, (John Wiley \& Sons, 2003). 

\bibitem{ade2016} P. A. R. Ade,  et al., `Planck 2015 results-xiii. cosmological parameters', Astronomy \& Astrophysics {\bf 594}, A13 (2016).

\bibitem{gnedin} N. Gnedin, `Cosmological Calculator for the Flat Universe', In: https://astro.uchicago.edu/$\sim$gnedin/cc/

\end{thebibliography}
\end{document}